\documentclass[preprint]{aastex6}

\usepackage{natbib}
\usepackage{xcolor}

\newcommand\species[2]{#1 {\sc #2}}

\def\eg{\mbox{e.g.}}

\def\teff{\mbox{T$_{\rm eff}$}}
\def\logg{\mbox{log~{\it g}}}
\def\vmicro{\mbox{$\xi_{\rm t}$}}
\def\kmsec{\mbox{km~s$^{\rm -1}$}}

\shorttitle{RRc Stellar Spectra}
\shortauthors{Sneden et al.}

\begin{document}

\title{The RRc stars: Chemical Abundances and Envelope Kinematics}

\author{
        Christopher Sneden\altaffilmark{1},
        George W.Preston\altaffilmark{2},
        Merieme Chadid\altaffilmark{3}, and
        Monika Adam{\' o}w\altaffilmark{1,4}
}

\altaffiltext{1}{Department of Astronomy and McDonald Observatory, 
                 The University of Texas, Austin, TX 78712, USA; 
                 chris@verdi.as.utexas.edu;madamow@astro.as.utexas.edu}
\altaffiltext{2}{Carnegie Observatories, 813 Santa Barbara Street, Pasadena, 
                 CA 91101, USA; gwp@obs.carnegiescience.edu}
\altaffiltext{3}{Universit\'e Nice Sophia--Antipolis, Observatoire de 
                 la C\^ote d’Azur, UMR 7293, Parc Valrose, F-06108, 
                 Nice Cedex 02, France; chadid@unice.fr}
\altaffiltext{4}{Toru{\'n} Centre for Astronomy, Faculty of Physics, 
                 Astronomy and Applied Informatics, Nicolaus Copernicus 
                 University in Toru{\'n}, Grudziadzka 5, 87-100 Toru\'n, 
                 Poland}

\begin{abstract}

We analyzed series of spectra obtained for twelve stable 
RRc stars observed with the echelle spectrograph of the du~Pont telescope 
at Las Campanas Observatory and we analyzed the spectra of RRc Blazhko stars 
discussed by \cite{govea14}.  
We derived model atmosphere parameters, [Fe/H] metallicities,
and [X/Fe] abundance ratios for 12 species of 9 elements.
We co-added all spectra obtained during the pulsation cycles 
to increase $S/N$ 
and demonstrate that these spectra give results superior to those obtained 
by co-addition in small phase intervals.  
The RRc abundances are in good agreement with those derived for the 
RRab stars of \cite{chadid17}. 
We used radial velocity measurements of metal lines and H$\alpha$ to construct 
variations of velocity with phase, and center-of-mass velocities.  
We used these to construct radial-velocity templates for use in 
low-medium resolution radial velocity surveys of RRc stars.  
Additionally, we calculated primary accelerations, radius variations, 
metal and H$\alpha$ velocity amplitudes, which we display as regressions 
against primary acceleration.  
We employ these results to compare the atmosphere structures of metal-poor 
RRc stars with their RRab counterparts.  
Finally, we use the radial velocity data for our Blazhko stars and the 
Blazhko periods of \cite{szczygiel07} to falsify the Blazhko oblique 
rotator hypothesis.

\end{abstract}


%%%%%%%%%%%%%%%%%%%%%%%%%%%%%%%%%%%%%%%%%%%%%%%%%%%%%%%%%%%%%%%%%%%%%%%%%%
\section{INTRODUCTION}\label{intro}
%%%%%%%%%%%%%%%%%%%%%%%%%%%%%%%%%%%%%%%%%%%%%%%%%%%%%%%%%%%%%%%%%%%%%%%%%%

RRc stars are core helium-burning first overtone pulsators that occupy the 
blue (hot) side of the RR Lyrae instability strip.  
They comprise approximately one quarter of the Galactic RR Lyrae population 
\citep{drake14, mateu12, soszynski14}.  
This estimate is compromised by the lower light amplitudes of RRc stars 
which tend to decrease frequency estimates, and confusion with WUMa-type 
contact binaries that tends to increase them.
Because they are hotter than the RRab, their metallic lines are weaker, 
so they were not included in the low resolution spectral abundance survey 
of \cite{layden94}, and they are under-represented in extant high-resolution 
spectroscopic investigations of the RR Lyrae stars (see \citealt{chadid17}
and references therein).  
The most extensive high resolution abundance study of RRc stars is that of 
\cite{govea14}.
Early estimates of RRc Luminosity via statistical parallax 
\citep{hawley86,strugnell86} derived from small samples (n $\sim$ 20) have 
recently been supplanted by the analysis of \cite{kollmeier13} based on a 
large RRc sample (n~=~242) provided by the All Sky Automated Survey\footnote{
available at http://www.astrouw.edu.pl/asas/?page=main}
(hereafter ASAS, \citealt{pojmanski02}).

The present RRc study is the sixth in a series devoted to pulsational 
velocities and chemical compositions of RR Lyrae stars from echelle spectra 
gathered with the Las Campanas Observatory (LCO) du~Pont telescope.   
The observations and reductions are nearly identical in our previous 
papers:  \cite{for11a,for11b}, \cite{chadid13}, 
\cite{govea14}[hereafter GGPS14], and \cite{chadid17}[hereafter CSP17]. 
This information will only be summarized here; see 
those studies for more detailed descriptions.

In this paper we present abundance analyses and radial velocity ($RV$) 
measurements of twelve stable RRc stars and the seven Blazhko\footnote{
The Blazhko effect is a slow (tens to hundreds of days) modulation of 
photometric and spectroscopic periods and amplitudes of some RR Lyrae
variables.  
It was first noticed in RW~Dra by \cite{blazhko07}.}
RRc stars investigated previously by GGPS14.
Our results permit comparisons of the abundances and pulsation properties 
of the RRc stars with those of the RRab sample of CSP17.
The basic observations, reductions, radial velocity determinations, 
spectrum stacking, and equivalent width measurements are presented in
\S\ref{obsred}, \ref{radial}, and \ref{phase}.
Model atmosphere, metallicity, and abundance analyses are discussed
in \S\ref{atmas}.
We explore RRc radial velocity variations in \S\ref{motionparams} and 
\S\ref{metalhalpha}.
Finally, we discuss failure of the rotational modulation hypothesis to 
explain the Blazhko effect in \S\ref{blazhko}.

%%%%%%%%%%%%%%%%%%%%%%%%%%%%%%%%%%%%%%%%%%%%%%%%%%%%%%%%%%%%%%%%%%%%%%%%%%
\section{OBSERVATIONS AND REDUCTIONS}\label{obsred}
%%%%%%%%%%%%%%%%%%%%%%%%%%%%%%%%%%%%%%%%%%%%%%%%%%%%%%%%%%%%%%%%%%%%%%%%%%

The du~Pont echelle spectrograph was employed with a 1.5 x 4.0\,arcsec 
aperture which yielded a resolving power
$R$~=~$\lambda/\Delta\lambda$ $\sim$~27000 at 5000~\AA.
The complete spectral domain was $\lambda\lambda$ 3400--9000, with small 
wavelength gaps in the CCD order coverage appearing at $\lambda$~7100~\AA\ 
and growing with increasing wavelength.
Table~\ref{tab-stars} contains basic data on these stars and those
studied by GGPS14.

Twelve stable RRc stars were chosen from the brightest RRc stars in the 
ASAS catalogue that were conveniently observable at LCO during our 
assigned nights.  
A total of 615 echelle observations were obtained during six nights in 
March 2014 and three nights in September 2014.  
Complete or almost complete phase coverage of the pulsations cycles was 
achieved for seven of the twelve stars.  
All twelve stars were observed at or near mid-declining light where line 
broadening is characteristically most favorable for spectroscopic analysis.

The pulsational periods of our program stars lie in the range
0.24~$\lesssim$~$P_{ASAS}$~$\lesssim$~0.38~days, with visual light 
amplitudes 0.23~$<$~$V_{amp;ASAS}$~$<$~0.55 (Table~\ref{tab-stars}).
In order to avoid deleterious effects of phase smearing, the maximum 
spectroscopic integration time was 400~s, with some 
observations as short as 200~s.
The maximum exposure times were thus less than 3\% of the shortest program
star period.
This led to negligible $RV$ variations during the individual stellar
observations.
The short exposures, apparent magnitudes the targets, and the modest du~Pont 
telescope aperture 
(2.5~m), led to low signal-to-noise in the extracted 
spectra, typically $S/N$~$<$~20.

Observations of the stars discussed previously by GGPS14 were 
obtained in 2009 and 2010, as circumstances permitted 
during our RRab survey (CSP17).
Their spectra were gathered in order to test the hypothesis that Blazhko 
periods are axial rotation periods, so the GGPS14 stars form a strongly 
biased Blazhko sample.  
Because of the relative rarity of Blazhko variables, the average visual 
magnitude of this sample at minimum light, $\langle V_{min}\rangle$~=~11.7, 
is more than one magnitude fainter than that of our stable RRc sample,
$\langle V_{min}\rangle$~=~10.5.
Therefore, of necessity, integration times were longer than those employed 
for the RRab stars in CSP17: the maximum was 1200~s, with typical 
values less than 900~s.  
Fortunately, the $RV$ amplitudes of RRc stars are smaller by a factor of
about two than those of RRab stars, so line broadening due to changing $RV$
during these longer integrations remained unimportant.  
Finally, because the Blazhko stars were observed at brief, randomly 
chosen times during the RRab survey, phase coverage for them is markedly
inferior to that obtained for our stable RRc stars.
We will discuss the axial rotations of the Blazhko sample
in \S\ref{blazhko}.
                                                              
We extracted spectra from the original CCD image files using procedures     
discussed previously \citep{preston00,for11a}; for the most part the        
description of these reduction steps need not be repeated here.  
However, as discussed in CSP17 we used one shortcut in scattered
light removal.
Our reduction steps do not explicitly account for scattered light, and
in \S3.1 of \cite{for11a} there is detailed consideration of its contribution
to the total flux in the extracted du~Pont echelle spectra.  
Fortunately, scattered light accounts for a near-uniform 10\% of the light 
in the spectral orders of interest to our work.
Therefore, like CSP17 we simply subtracted the 10\% and renormalized to 
get the final spectra ready for analysis.

%%%%%%%%%%%%%%%%%%%%%%%%%%%%%%%%%%%%%%%%%%%%%%%%%%%%%%%%%%%%%%%%%%%%%%%%%%
\section{RADIAL VELOCITIES}\label{radial}
%%%%%%%%%%%%%%%%%%%%%%%%%%%%%%%%%%%%%%%%%%%%%%%%%%%%%%%%%%%%%%%%%%%%%%%%%%

We accumulated $\sim$50$\pm$15 spectra for each stable RRc.  
These allow us to examine $RV$ variations of the metallic 
lines and H$\alpha$ with pulsation phase and to compare the relationship 
between velocity amplitude and light amplitude of first overtone 
pulsators with that established for the RRab fundamental pulsators.  
Likewise, we can compare the envelope structures revealed by the different 
$RV$ variations of the near-photospheric metal line velocities and 
the velocities at the very small continuum optical depths high in the 
atmosphere where the core of H$\alpha$ is formed.

As done for the RRab sample, we flattened and stitched together thirteen 
echelle orders of each spectrum spanning the wavelength interval  
$\lambda\lambda$4000$-$4600~\AA\ by use of an IRAF script prepared by Ian 
Thompson (private communication).  
This is the same spectral interval used in our previous papers.  
Flattened, normalized spectra of the echelle order containing H$\alpha$ near 
its center were used to measure H$\alpha$ radial velocities.  
As in our previous investigations we measured $RV$s with the $IRAF/fxcor$ 
package, using the same reference spectra for metal lines (star CS~22874-009) 
and for H$\alpha$ (CS~22892-052), and the same primary 
$RV$ standard (HD~140283). 
We used the same Gaussian fitting function to locate line centers 
as discussed in detail by CSP17, and the 
same procedure for measurement of occasional asymmetric H$\alpha$ cores. 
That is, if an H$\alpha$ core is asymmetric we always fit a Gaussian to 
the flux minimum and the steepest wing of the profiles.
Thus, our $RV$s for RRc stars are directly comparable to those reported 
previously for the RRab stars by CSP17.
The derived $RV$s for metallic lines and H$\alpha$ are given in 
Table~\ref{tab-rvobs}.

\begin{figure}
\epsscale{1.00}
\plotone{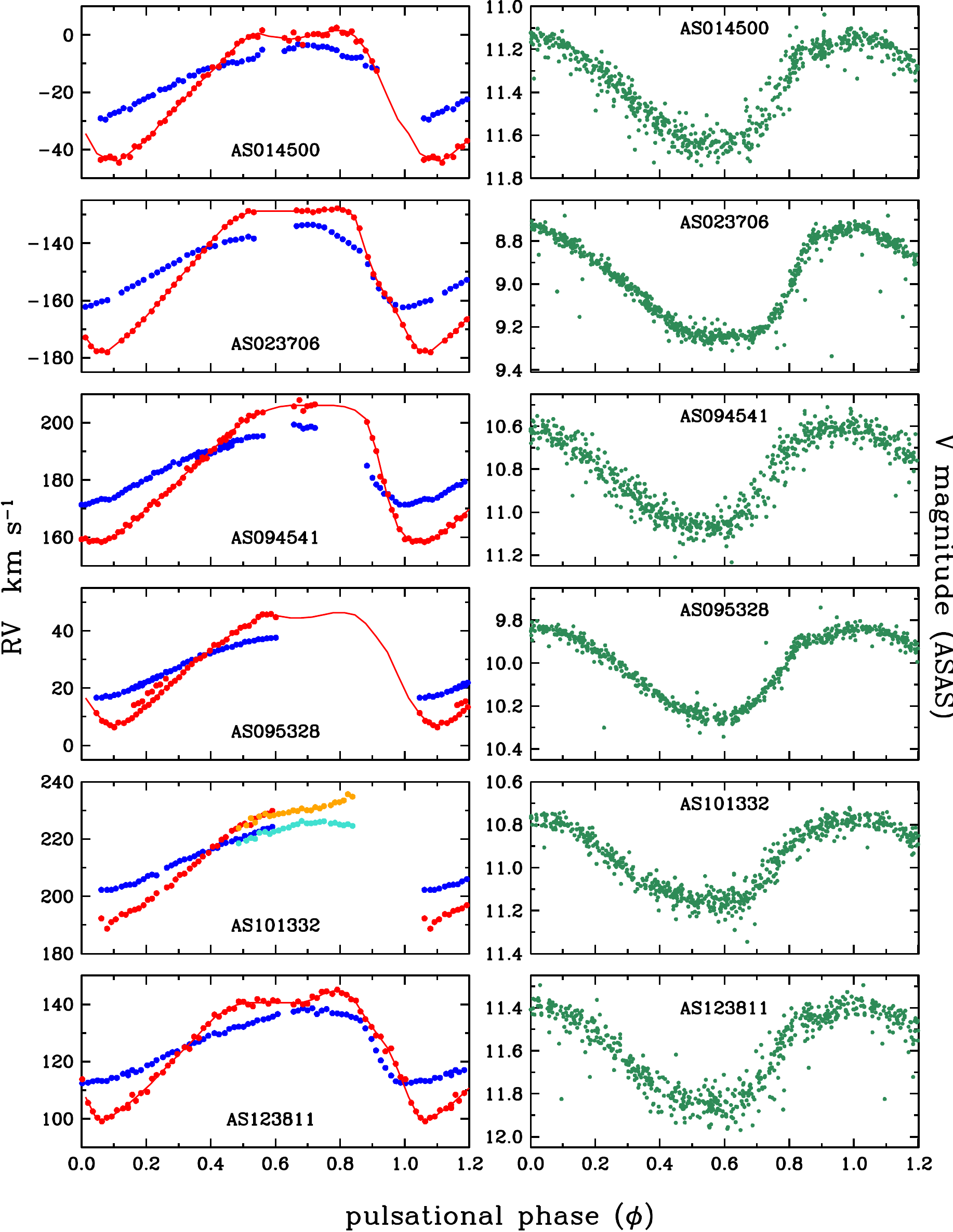}
\caption{\label{f1}
\footnotesize
   Left-hand panels:  Derived $RV$s for metal lines (blue points) and
   H$\alpha$ (red points and mean line) as functions of pulsational
   phase $\phi$ for the first six of the stable RRc stars newly observed
   for this study.
   the $RV$ extent of each panel is 60~\kmsec, approximately centered on
   the stellar systemic velocity.
   Right-hand panels: ASAS database $V$ magnitudes for these stars.
   The $P$ and $T_0$ values to produce these light curves are those
   derived from the $RV$ analyses of this study.
   For AS101332 the $RV$ data do not match well even though the observations
   were obtained only two nights apart (Table~\ref{tab-rvobs}).
   We show this by using turquois color for the metal line and orange
   color for the H$\alpha$ points in this star's $RV$ plot.
}
\end{figure}

\begin{figure}
\epsscale{1.00}
\plotone{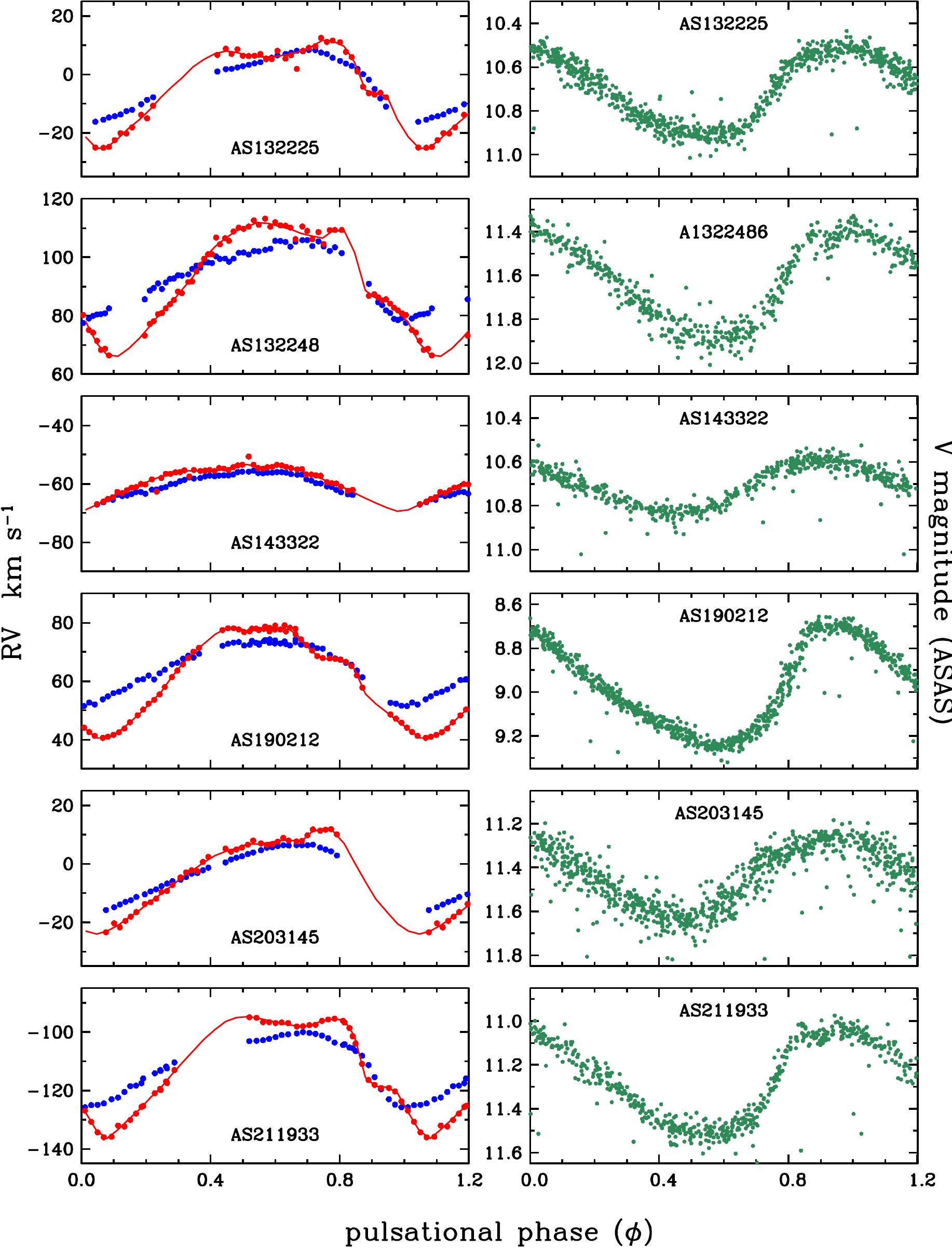}
\caption{\label{f2}
\footnotesize
   Velocities and $V$ magnitudes as functions of pulsational phase $\phi$
   for the second set of six stable RRc stars of this study.
   All points and lines are as in Figure~\ref{f1}.
}
\end{figure}

Plots of $RV$ versus pulsation phase were computed for the twelve 
stable RRc stars by use of ASAS periods and initial Heliocentric 
Julian Days listed in Table~\ref{tab-stars}.  
These curves are shown in the left-hand panels of Figures~\ref{f1} 
and \ref{f2}. 
The $RV$ curves for the GGPS14 stars are shown in their paper.

We adopted zero points of phase for our radial velocity curves and ASAS 
visual light curves by requiring that the minima of visual magnitudes 
and radial velocities be coincident.  
Such coincidence, first noticed by \cite{sanford30} for two classical 
cepheids, is verified for RR Lyrae stars by data in Table~\ref{tab-coin2} 
that we constructed from examination of near-contemporaneous radial 
velocity and visual photometric observations made by several groups for 
applications of the Baade-Wesselink method.  
A definitive investigation of coincidence among the RR Lyrae stars of M3 
(not attempted here) could be conducted by use of the recent elegant 
observations of \cite{jurcsik15,jurcsik17}. 
Table~\ref{tab-coin2} provides arguable evidence for small offsets 
($\simeq$0.01P) for both RRab and RRc stars, but these do not exceed 
their standard deviations.  
We are obliged to assume coincidence because, unfortunately, the ASAS 
photometric ephemerides of our stars do not predict our observed times 
of minimum $RV$. 
Our spectroscopic observations, all made near JD 2456800~$\pm$~100, were 
obtained some 1600 days or 5000 pulsation cycles after the most 
recent published ASAS photometry (through Dec. 31, 2009), so we suspect 
unidentified problems with the published light curve ephemerides. 
The ASAS release of photometric data from 2010 forward is a work in 
progress, so we can expect resolution of this ephemeris problem in 
the near future. 
In the present study we arbitrarily adopted values of $HJD0$ that 
produced visual light maxima near phases 0.00, consistent with the 
results in Table~\ref{tab-coin2}. 
We also made small adjustments to the ASAS period values to produce light
curves with the most internal consistency.  
We show these light curves in the right-hand panels of Figures~\ref{f1} and
\ref{f2}.
See GGPS14 for light curve plots of the Blazhko stars.

Note that our radial velocity observations of AS101332 made on 
JD~2456740 and 2456742 do not superpose well; see Figure~\ref{f1}.  
Were this to be a Blazhko phenomenon the Blazhko period would have to be 
short and the  amplitude of the phase variation would have to be large.  
We see no evidence of this in the ASAS photometry, and the star was not 
identified as a Blazhko variable by \cite{szczygiel07}.  
We offer no other explanation for the behavior of this star.

The eight stable RRc stars (those in column~7 of 
Table~\ref{tab-stars} for 
which we can estimate $RV$ amplitudes) comprise a homogeneous group with 
respect to radial velocity and visual light amplitudes (24.2~$\pm$~5.1~\kmsec,
and 0.46~$\pm$~0.09~mag., respectively). 
If we omit the low amplitude outlier AS143322, the standard deviations 
are even smaller ($\pm$2.3~\kmsec\ and $\pm$0.05~mag, respectively).  
We used our estimates of the extrema defined by the data in 
Figures~\ref{f1} and \ref{f2} to calculate the visual light and radial 
velocity amplitudes listed in columns~7 and 8 of Table~\ref{tab-stars};
these will be discussed in \S\ref{motionparams}.

Our small \logg~=~2.3 for AS143322 rules out classification as an
SX~Phe star like those studied \cite{nemec17}. 
The small [Fe/H]~=~$-$1.48 of this star rules out identification as a 
$\delta$~Scuti star, so perforce we retain AS143322 as an unusual 
short-period, low-amplitude RRc (Table~\ref{tab-stars}).

%%%%%%%%%%%%%%%%%%%%%%%%%%%%%%%%%%%%%%%%%%%%%%%%%%%%%%%%%%%%%%%%%%%%%%%%%%
\section{PHASE CO-ADDITION AND EQUIVALENT WIDTHS}\label{phase}
%%%%%%%%%%%%%%%%%%%%%%%%%%%%%%%%%%%%%%%%%%%%%%%%%%%%%%%%%%%%%%%%%%%%%%%%%%

To perform model atmosphere and chemical composition analyses we again
re-stitched and flattened all of the echelle orders in each observed
spectrum, taking care in these procedures to excise anomalous cosmic rays 
and other radiation events, to produce 
continuum-normalized spectra ready for equivalent width measurements.
The spectra also were corrected to zero velocity using the instantaneous 
velocity shift for an observation that was 
derived as part of the computations discussed in \S\ref{radial}.
However, the combination of RRc line strengths and the constraints
of our observational parameters necessitated further manipulations to
produced useful spectra for atmospheric analysis.

\begin{figure}
\epsscale{0.75}
\plotone{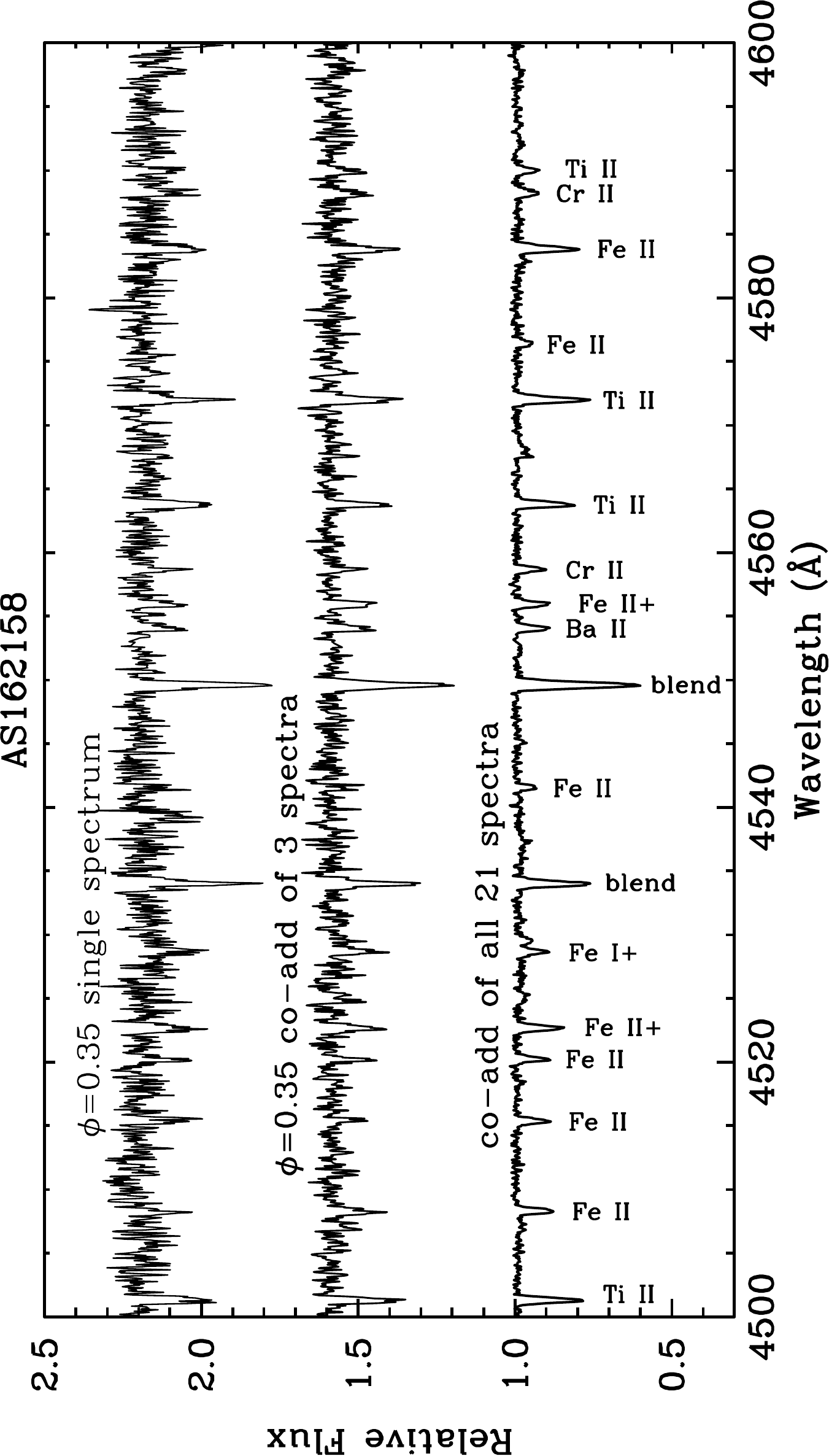}
\caption{\label{f3}
\footnotesize
   An example of co-addition of spectra, using the very metal-poor star
   AS162158.
   The chosen spectral region is rich in the lines of iron-group ions, and
   contains the well-known  resonance line of \species{Ba}{ii} at 4554~\AA.
   The top spectrum is one of the individual observed spectra obtained at
   pulsational phase $\phi$~$\simeq$~0.35.
   The middle spectrum is the co-addition of three spectra near this
   phase; this spectrum was used in the phase-based atmospheric analysis
   for AS162158.
   The bottom spectrum is the mean spectrum of all 21 spectra gathered for
   this star.
   Phase information has been sacrificed to substantially increase the
   $S/N$ of this co-added spectrum.
}
\end{figure}

Average RRc parameters reported by GGPS14 were \teff~$\sim$~7200~K, 
\logg~$\sim$~2.3, and [Fe/H]~$\sim$~$-$2.0.  
Such very warm and metal-poor giant stars have weak-lined spectra.
The modest $S/N$ of our individual short-exposure spectra of RRc
stars were  ill-suited to derivation of model atmosphere quantities
and abundances of individual elements.
The most reliable stellar abundances usually arise from equivalent width 
($EW$) measurements of weak absorption features that lie on the linear 
part of the curve-of-growth, that is lines that have reduced widths
log($RW$)~=~log($EW/\lambda$)~$<$~$-$5.0 (\eg, 45~m\AA\ at 
$\lambda$~=~4500~\AA).
Such lines are sometimes detectable on our individual spectra but their 
measured $EW$'s have large uncertainties.
We illustrate this problem in Figure~\ref{f3}.
The top spectrum in this figure is one integration of the star 
AS162158, one of the most metal-poor stars of our sample.
Many absorption lines of (mostly) Fe-group ionized species are in this 
spectral region, but nearly all of them have $EW$~$\lesssim$~80~m\AA, 
or log($RW$)~$\lesssim$~$-$4.7.
It is clear from inspection of this Figure~\ref{f3} 
``snapshot'' spectrum of AS162158 that $EW$s of weak detected 
lines are very uncertain.
Figure~9 of GGPS14 shows another example, that of the more metal-rich
AS230659, in the 5200~\AA\ spectral region where the stellar fluxes are
higher.
Even in this case $EW$ values for weak lines are not easy to determine
on individual spectra.

Therefore, following the technique developed by \cite{for11b}, we co-added
individual spectra in phase bins no wider than $\delta\phi$~=~0.1, producing
higher $S/N$ spectra representative of several phase intervals throughout 
our program star pulsational cycles.
In Table~\ref{tab-phase} we list all phase bins for each program star, the
mean phase of the bins and the number of individual spectra contributing 
to the mean spectra.
An example of the spectrum phase averaging is given in the middle spectrum
of Figure~\ref{f3}; the increase in $S/N$ by the co-addition of three 
AS162158 spectra obtained near $\phi$~=~0.35 is apparent.
Hereafter these will be called ``phase-mean'' spectra for each star.

Even these co-additions were insufficient to reveal very weak features.
This did not severely affect our analyses of species with many transitions
on our spectra, such as \species{Fe}{i}, \species{Fe}{ii}, and 
\species{Ti}{ii}.
But many species of interest, \eg, \species{Cr}{i} or \species{Ba}{ii},
are represented by only very few lines; these species often cannot be
studied in our more metal-poor RRc stars, even with phase co-added spectra.
Fortunately RRc stars have relatively mild atmospheric physical state changes 
during their pulsation cycles.
The mean ranges of temperatures and gravities of the program stars 
to be discussed in \S\ref{atmas} are
$\langle\teff_{,max} - \teff_{,min}\rangle$~= 540~$\pm$~250~K and
$\langle\logg_{,max} - \logg_{,min}\rangle$~= 0.7~$\pm$~0.5 .
Therefore we also performed co-additions of all spectra for each program star.
This procedure of course threw away all pulsational phase information,
in return for greatly increased $S/N$.
These spectra were then analyzed for stellar parameters and abundance ratios.
The bottom spectrum in Figure~\ref{f3} shows the result of 
co-adding all 21 spectra of AS162158.
Hereafter these will be called ``total'' spectra for each star.

We performed $EW$ measurements with the recently-developed 
$pyEW$ code\footnote{
https://github.com/madamow}
This $EW$ package is a Python script that, for a given stellar absorption 
line, fits a multigaussian function to a fragment of the spectrum centered 
on the line under examination.
The multigaussian approach allows one to handle complex features that
included in blending contaminants.
The $pyEW$ code can be used in a completely automatic way, or it can be run 
in interactive mode which allows to user to modify a solution that was found 
automatically.
It can be used in a hybrid fashion in which some of the lines are analyzed
automatically and others are subjected to user scrutiny.

%%%%%%%%%%%%%%%%%%%%%%%%%%%%%%%%%%%%%%%%%%%%%%%%%%%%%%%%%%%%%%%%%%%%%%%%%%
\section{Atmosphere and Abundance Derivations from the Spectra}\label{atmas}
%%%%%%%%%%%%%%%%%%%%%%%%%%%%%%%%%%%%%%%%%%%%%%%%%%%%%%%%%%%%%%%%%%%%%%%%%%

We determined atmospheric parameters (\teff, \logg, \vmicro, [Fe/H]) and
ratios [X/Fe] for our RRc stars with essentially identical procedures to
those used by CSP16 for their sample of RRab stars.
Here we summarize the CSP17 procedures, mostly as described in their \S3.1.
At the end of this section we comment on the metallicity range of our RRc
stars.

\subsection{Model Atmosphere Parameters}\label{model}

Model atmospheres were interpolated in the \cite{kurucz11} ATLAS grid\footnote{
http://kurucz.harvard.edu/grids.html.}
and we employed a new version of the local thermodynamic equilibrium
(LTE) line analysis program $MOOG$ \citep{sneden73}\footnote{
Available at http://www.as.utexas.edu/~chris/moog.html.}
to derive abundances for individual lines.
The $pyMOOGi$ code \footnote{ 
https://github.com/madamow}
 is a Python ``wrapper'' that retains the basic synthetic spectrum 
computations of $MOOG$ but introduces improved interactive graphical 
capabilities through standard Python library functions.
Currently $pyMOOGi$ is available for the {\it synth} and {\it abfind} $MOOG$
drivers that we need in this study. 
We adopted the CSP17 atomic line list without change.

The input model atmospheres were iteratively changed until the ensemble of
\species{Fe}{i} and \species{Fe}{ii} 
lines showed no elemental abundance trends
with excitation energy $\chi$, line strength log($RW)$, and ionization state,
and the input ``$\alpha$-enhanced'' model metallicity matched the 
derived [Fe/H] value.
We generally discarded very strong lines in our abundance analyses, those 
with log($RW$)~$>$~$-$4.5 ($EW$~$\simeq$~140~m\AA\ at 4500~\AA).
The strengths of those lines increase sharply with increasing log($RW$),
clearly reflecting outer atmosphere conditions in pulsating RRc stars
that are not well described by our simple plane-parallel atmosphere and
LTE line analysis assumptions.
The derived model atmosphere and metallicity values for all stars in all
co-added phase bins are given in Table~\ref{tab-phase}.  
These quantities were then averaged on a star-by-star basis to form 
phase-means values for each of the 19 program stars.
These are entered in Table~\ref{tab-phasemean}.

For each star we also performed the model atmospheric analysis using the
total spectra.  
In Table~\ref{tab-total} we list the results.
Abundances of other elements were derived from each phase-mean spectrum and
each total spectrum, but we deem only those from the total spectra to be
reliable enough for further study.  

\begin{figure}
\epsscale{1.00}
\plotone{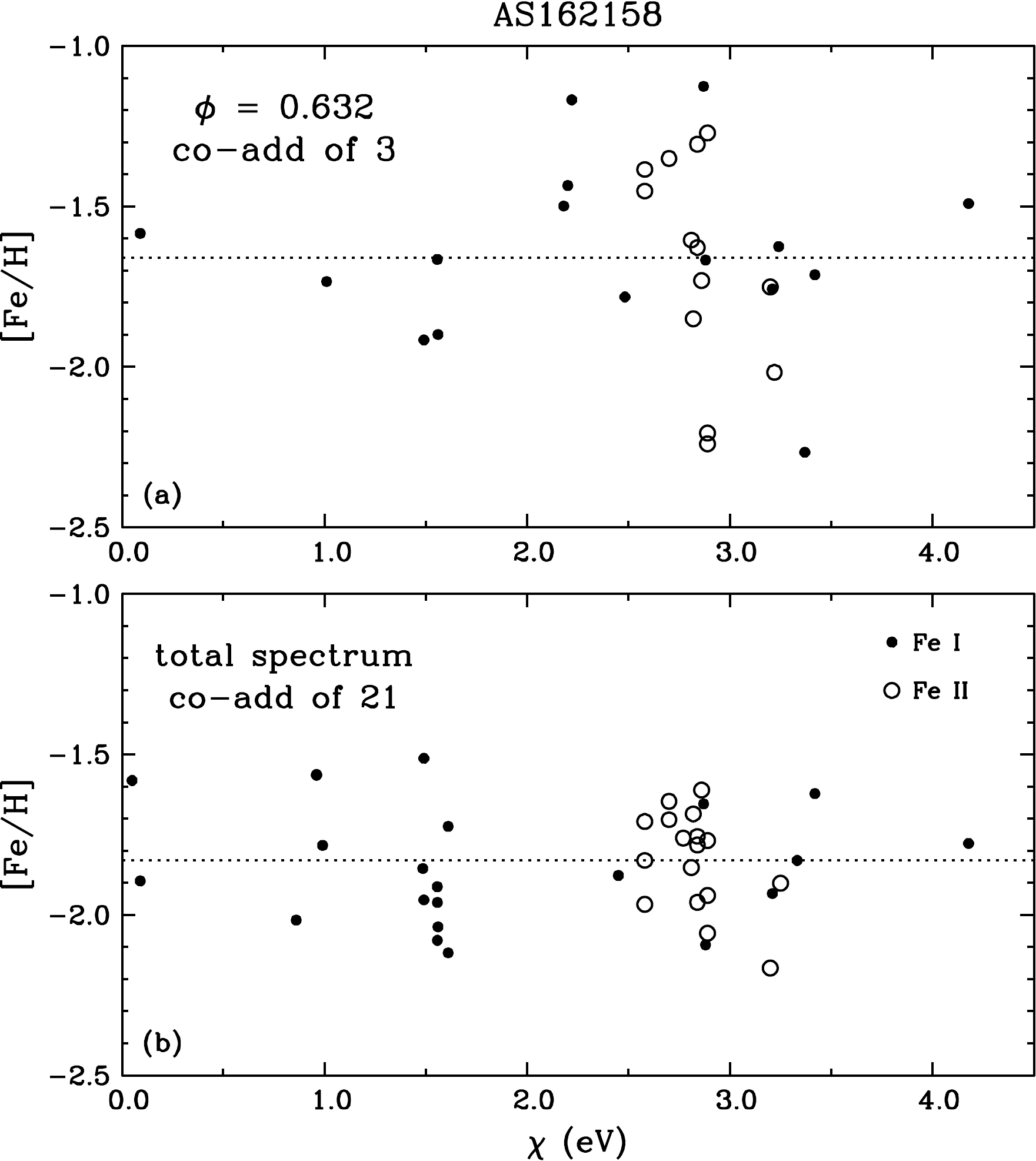}
\caption{\label{f4}
\footnotesize
   Correlations of excitation potentials $\chi$ with derived line 
   abundances of \species{Fe}{i} (filled circles) and \species{Fe}{ii} 
   (open circles) line abundances in AS162158.
   Panel (a) has the abundances from the co-addition of three spectra at phase 
   $\langle\phi\rangle$~=~0.632 are shown, and panel (b) has the abundances
   from co-addition of all 21 AS162158.
   The dotted lines in each panel represent the mean abundance of Fe from
   the species abundances in that panel.
}
\end{figure}

The entries in Table~\ref{tab-total} can compared to the phase-mean 
values of Table~\ref{tab-phasemean}.
Note especially the increase in the number of transitions available for
determining the abundances from \species{Fe}{i} and \species{Fe}{ii}
lines.
As an example of the improvement in ability to measure useful spectral
lines by averaging all observations of a star to produce a total spectrum, 
we show in Figure~\ref{f4} the line-by-line abundances 
for Fe in AS162158.
In panel (a) we have selected the results for phase $\phi$~=0.632,
which is based on co-addition of 3 individual observations, and
in panel (b) the results of the total spectrum with co-addition of
all 21 observations are shown.
The decrease in line-to-line scatter in the analysis of the total spectrum
is clear by visual inspection, and confirmed by the statistics of
Tables~\ref{tab-phase} and \ref{tab-total}.
From the $\phi$~=~0.632 spectrum analysis 
$\sigma$(\species{Fe}{i})~=~0.28 from 16 lines and
$\sigma$(\species{Fe}{ii})~=~0.33 from 13 lines, while from the total 
spectrum analysis
$\sigma$(\species{Fe}{i})~=~0.19 from 22 lines and
$\sigma$(\species{Fe}{ii})~=~0.15 from 17 lines.

\begin{figure}
\epsscale{1.00}
\plotone{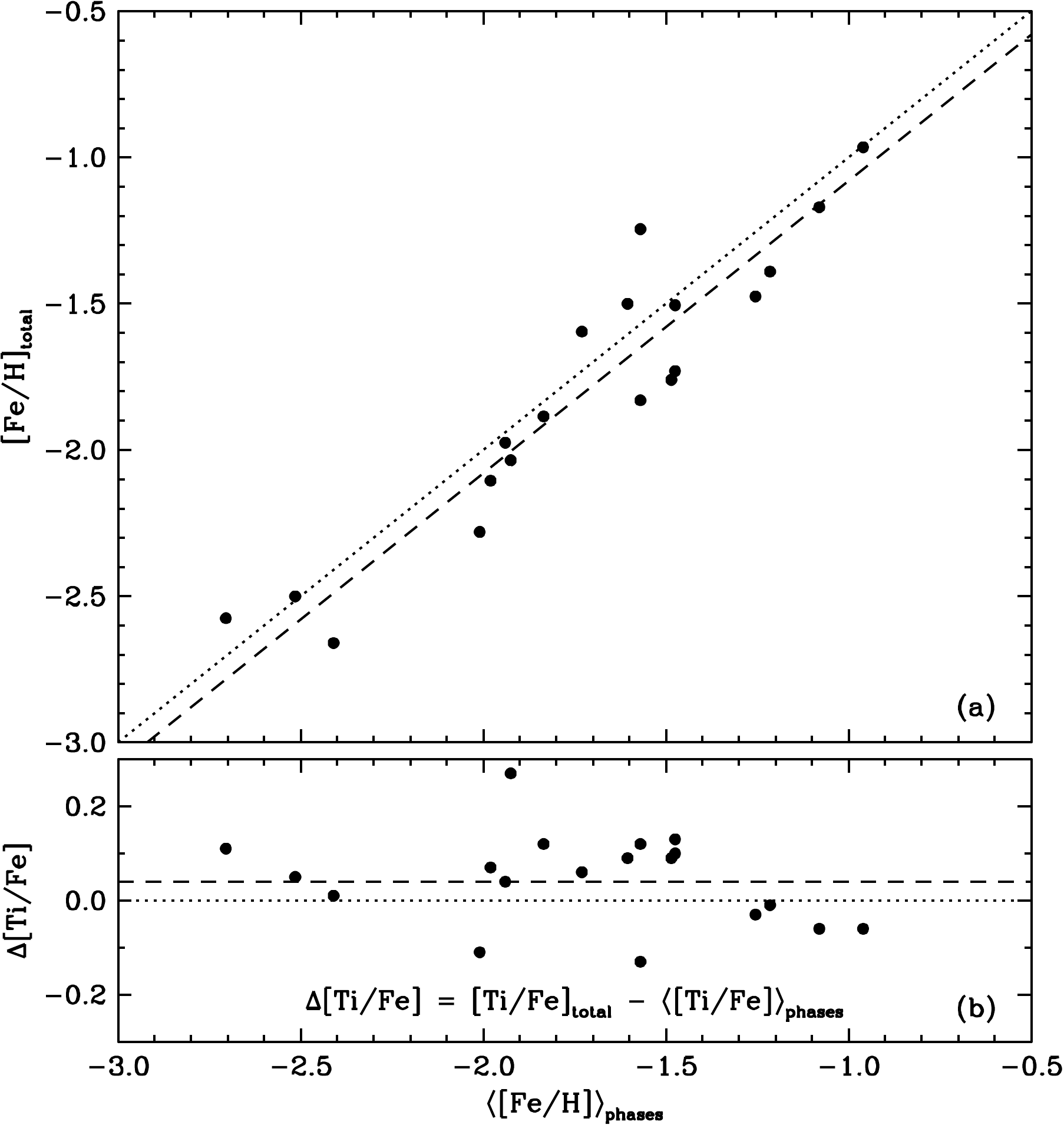}
\caption{\label{f5}
\footnotesize
   Panel (a): Comparison of metallicity values computed by forming for each 
   star the mean of values derived at the co-added phase points (labeled 
   $\langle$[Fe/H]$\rangle$  $-$ all phases), and the value derived from
   the total spectrum formed from the addition of all individual 
   spectra regardless of their phases (called [Fe/H] $-$ mean spectrum).
   The dotted line shows equality of the metallicities, and the
   dashed line shows the derived mean offset in these values.
   Panel (b): Differences between the relative abundance ratios [Ti/Fe]
   between those determined from the total spectra and those determined
   from the averages of individual phases.
   An example of co-addition of spectra, using the very metal-poor star
   AS162158.
   The dotted line represents equality of the two [Ti/Fe] values,
   and the dashed line represents the derived mean offset from equality.
}
\end{figure}

The metallicities of phase-mean and total spectra correlate well, as
can be seen in panel (a) of Figure~\ref{f5}.
The mean metallicity offset, 
$\langle$[Fe/H]$_{total}-$[Fe/H]$_{phase mean}\rangle$ = $-$0.08~$\pm$~0.04
($\sigma$~=~0.17, 19 stars).
The small offset lies well within the star-to-star scatter.  
If the offset is real then the higher $S/N$ of the total spectrum might be 
the cause, as the continuum setting becomes easier as the $S/N$ increases.
However, the total spectrum is an average over all phases of an ever-changing
RRc target, and thus smears the phase-specific information that is 
available in the phase-mean spectra of a star.  
Further investigation of this small offset is beyond the scope of our
work, and it is clear that all of our methods of deriving RRc [Fe/H] values
yield essentially the same answers over a 2-dex metallicity range.

Program star AS190212 deserves comment for several reasons.
\textit{(1):} It is one of the most metal-poor stars of our sample.
\textit{(2):} From its total spectrum we derive an unusually
high surface gravity (\logg~= 3.7; Table~\ref{tab-total}).
\textit{(3):} Its H$\alpha$ $RV$ curve has a ``standstill'' flat portion
in the phase range $\phi$~$\simeq$~0.70$-$0.85 (Figure~\ref{f2}) that is
not obvious in any of our other RRc stars.
AS190212 under its variable star name of MT~Tel has several literature
metallicity estimates: [Fe/H]~$\sim$~$-$2.5 (\citealt{przybylski83}, from a
``coarse analysis'' with respect to a Hyades cluster star);
[Fe/H]~= $-$1.63 (\citealt{solano97}, based on two \species{Fe}{ii} lines);
and [Fe/H]~= $-$1.85 (\citealt{fernley98}, using a $\Delta S$ calibration from
\citealt{fernley97}).
No previous extensive high-resolution spectroscopic analysis of AS190212
has been published.
Our analysis of this star was conducted in an identical manner to the others
of our sample, and the derived metallicities from \species{Fe}{i} and
\species{Fe}{ii} lines of the phase-averaged spectra
(Table~\ref{tab-phasemean}) agree well with those of the total spectra
(Table~\ref{tab-total}).
The gravity value was derived from a Saha equilibrium balance between
20 \species{Fe}{i} and 10-13 \species{Fe}{ii} lines.
Additionally, the gravity we derive from the AS190212 is not unusually high:
\logg~=~3.2.

The relatively large gravity of AS190212 places it not far from 
the blue metal-poor (BMP) main sequence studied by \cite{preston00}.
Their atmospheric analyses of Du Pont spectra of nearly 50 slowly rotating
BMP stars yielded $\langle \logg\rangle$~$\simeq$~4.3, with only three of their
stars having \logg~$\leq$~3.9.
The atmospheric parameters of AS190212 are similar to those of metal-poor
SX~Phoenecis stars that exist in small numbers in the halo field
(\eg, \citealt{preston99}) and in globular clusters (\eg, M53,
\citealt{nemec95}).
However, SX~Phe variables typically have very short periods,
$P$~$\sim$~0.04~days, while the period of AS190212 is 0.32~days.
This probably negates association of this star with the SX~Phe class.
Another intriguing possibility is that AS190212 might be like
the mass-transfer binary star OGLE-BLG-RRLYR-02792, whose brighter
primary has a very low mass ($M$~=~0.26$M_\odot$) star that exhibits
RR~Lyrae photometric variations \citep{pietrzynski12}.
Those authors suggest that the complex history of this binary has placed it
for a short time in the instability strip domain of the RRc stars.
However, while the OGLE primary has about the right temperature,
\teff~$\simeq$~7300~K, its pulsation period is 0.63~days, about double
the periods of our RRc sample.
AS190212 most likely remains a true RRc star, but one should keep in mind
the difficulty in assigning variability classes to stars in this
\teff-\logg\ domain.

\subsection{Elemental Abundances}\label{relabunds}

For the total spectra we also derived abundance ratios [X/Fe] from 12 species
of 9 elements.
The star-by-star values and overall species means are listed in 
Tables~\ref{tab-alphas} and \ref{tab-fegroup}.
To compute these [X/Fe] numbers we adopted the recommended solar photospheric
abundances of \cite{asplund09}, and used the same species in numerator
and denominator of the fractions, \eg\ the [Ti/Fe]$_{\rm I}$ value was 
determined from the [Ti/H]$_{\rm I}$ and [Fe/H]$_{\rm I}$ abundances, and 
[Ti/Fe]$_{\rm II}$ was from [Ti/H]$_{\rm II}$ and [Fe/H]$_{\rm II}$ abundances.
Many of species in these tables are represented by only a handful of
lines, only one in the case of \species{Si}{i}.
Obviously the abundances of these species should be viewed with caution.

We derived abundances from six species of four $\alpha$ and $\alpha$-like 
elements; these are listed in Table~\ref{tab-alphas}.
There are enough \species{Ti}{ii} lines to make a comparison between the
phase-mean and the total spectra.  
In panel (b) of Figure~\ref{f5} we show the [Ti/Fe] differences between
these two abundance approaches, and they are obviously small.
As defined in the figure legend, $\langle\Delta$[Ti/Fe]$\rangle$~= 
$+$0.04~$\pm$~0.02 ($\sigma$~=~0.10, 19 stars).
As with the metallicity comparison, the total and phase-mean spectra 
yield the same values for Ti/Fe ratios.

\begin{figure}
\epsscale{1.00}
\plotone{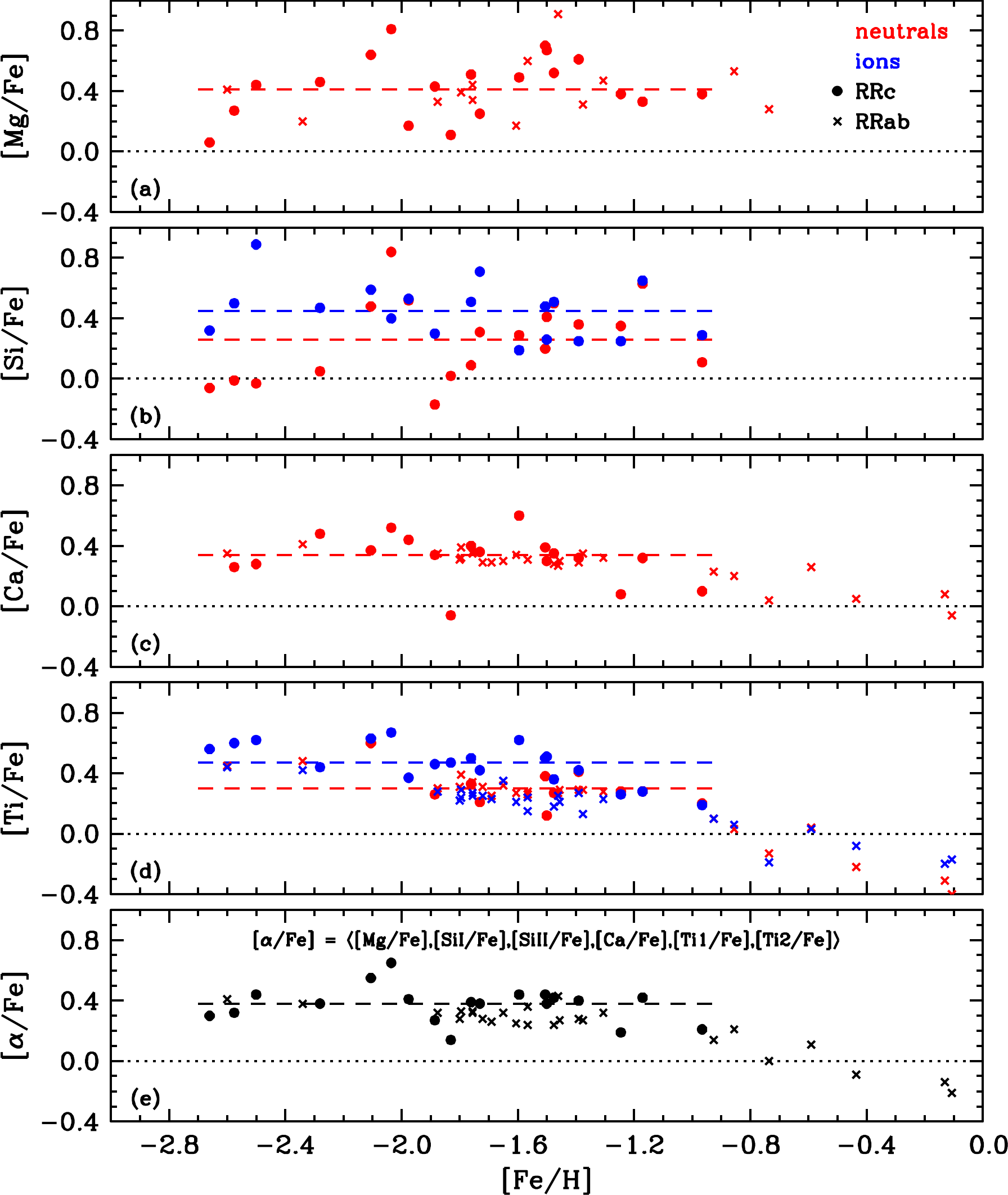}
\caption{\label{f6}
\footnotesize
   Relative abundance ratios [X/Fe] for the $\alpha$ and $\alpha$-like
   elements.
   The colors and point types are defined in the legend of panel (a). 
   The RRc stars are results from the present study, and the RRab stars
   are those of CSP17.
   The dished lines in each panel are the overall species mean [X/Fe] values.
   In panel (e) we show unweighted means of the species abundance
   from panels (a)-(d), as defined in the panel (e) legend.
}
\end{figure}

In Figure~\ref{f6} we plot our $\alpha$-element abundances 
and those of RRab stars studied by CSP17.
The line-to-line and star-to-star scatters are larger for Mg (panel a)
and Si (panel b) than they are for Ca (panel c) and Ti (panel d).
Mg is represented only by a few \species{Mg}{i} lines that are often 
strong and thus affected by microturbulent velocity choices.
\species{Si}{i} has only one routinely detectable transition at 3905~\AA.
This transition is known to yield unreliable Si abundances 
(\eg, \citealt{sneden08b}).
Several \species{Si}{ii} lines are available for analysis, but it is 
difficult to calibrate them because they usually are unreported in the 
vast majority of metal-poor (typically cooler) stars.
In panel (e) we show a simple straight mean for all the $\alpha$ species
abundances.
These values are essentially invariant across our entire RRc metallicity domain.

\begin{figure}
\epsscale{1.00}
\plotone{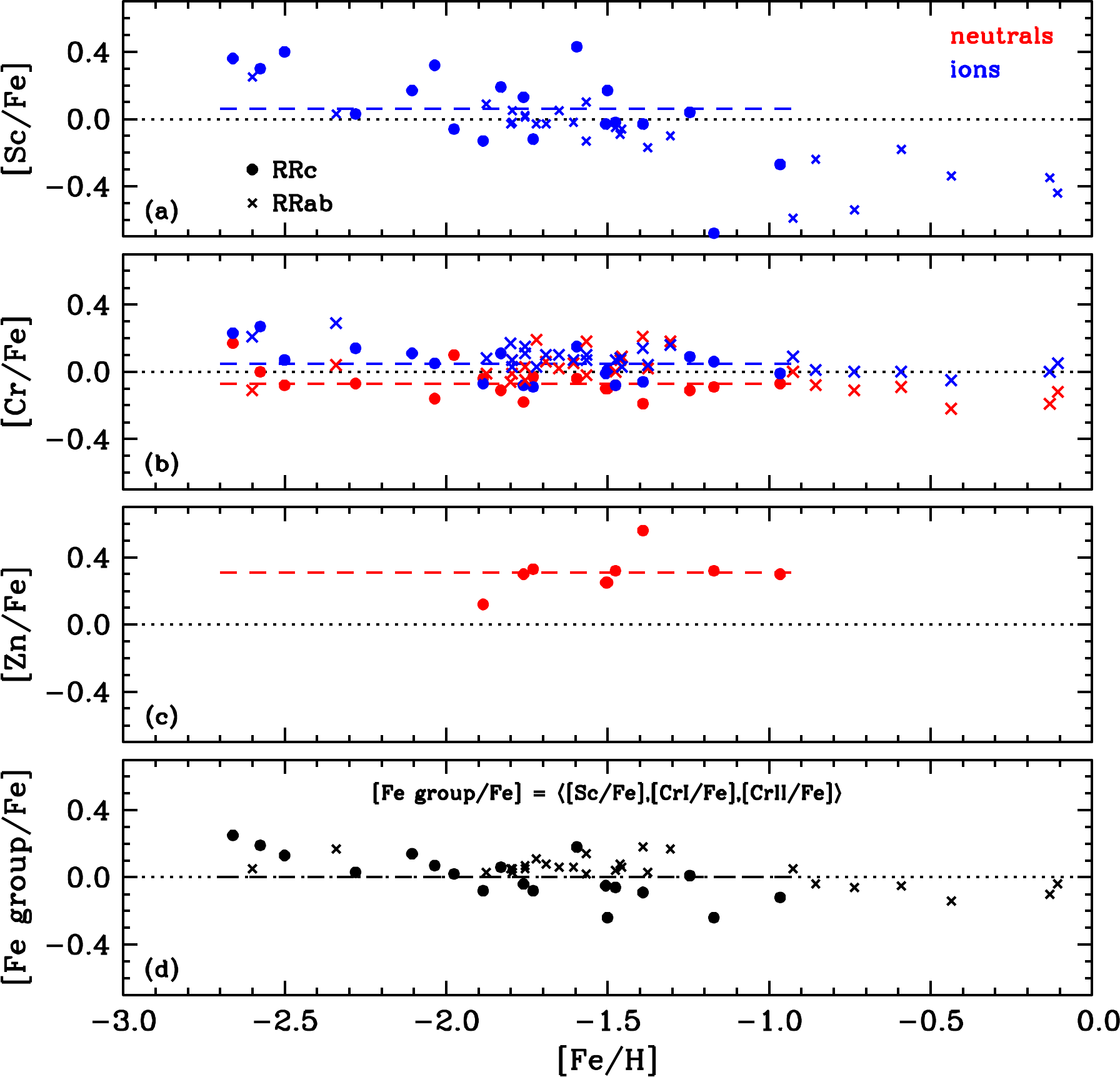}
\caption{\label{f7}
\footnotesize
   Relative abundance ratios [X/Fe] for Fe group elements.
   The colors, symbols, and lines are as in Figure~\ref{f6}.
   In panel (d) we show unweighted means of the species abundances
   from panels (a) and (b) only; the uncertain [Zn/Fe] abundances
   of panel (3) are not included.
}
\end{figure}

We also derived abundances from four species of three Fe-group elements
beyond Fe itself.
These are listed in Table~\ref{tab-fegroup} and displayed in 
Figure~\ref{f7}.  
Abundances deduced from the 1-2 \species{Zn}{i} lines (panel c) are few
in number and we do not consider them 
to be reliable.
The Fe-group means excluding Zn are shown in panel (d) of this figure; they 
clearly have small star-to-star variations and are consistent with 
[Fe group/Fe]~$\simeq$~0 over the whole metallicity regime.

Finally, the only $n$-capture elements with reliable abundances derivable 
over the whole metallicity range are Sr and Ba.
These elements have the only strong transitions beyond the near-UV: 
\species{Sr}{ii} 4077 and 4215~\AA, and \species{Ba}{ii} 4554, 5853, 6141, 
and 6496~\AA.
However, the \species{Sr}{ii} lines are often very saturated or otherwise
compromised in their complex spectral regions, so we only quote Sr abundances
for 12 of the 19 program stars in Table~\ref{tab-fegroup}.
The star-to-star scatter for [Sr/Fe] is large, $\sigma$~=~0.50).
Caution is warranted in interpretation of our results for both Sr and Ba, 
but there is little evidence for departures from 
[Sr,Ba/Fe]~=~0 in our data;
more detailed $n$-capture analyses for RRc stars would be welcome in the future.

%%%%%%%%%%%%%%%%%%%%%%%%%%%%%%%%%%%%%%%%%%%%%%%%%%%%%%%%%%%%%%%%%%%%%%%%%%  
\section{PHOTOSPHERIC MOTION PARAMETERS}\label{motionparams}   
%%%%%%%%%%%%%%%%%%%%%%%%%%%%%%%%%%%%%%%%%%%%%%%%%%%%%%%%%%%%%%%%%%%%%%%%%%  

%%%%%%%%%%%%%%%%%%%%%%%%%%%%%%%%%%%%%%%%%%%%%%%%%%%%%%%%%%%%%%%%%%%%%%%%%%  
\subsection{Amplitude Diagrams}\label{amplitudes}
%%%%%%%%%%%%%%%%%%%%%%%%%%%%%%%%%%%%%%%%%%%%%%%%%%%%%%%%%%%%%%%%%%%%%%%%%%  

\begin{figure}
\epsscale{1.00}
\plotone{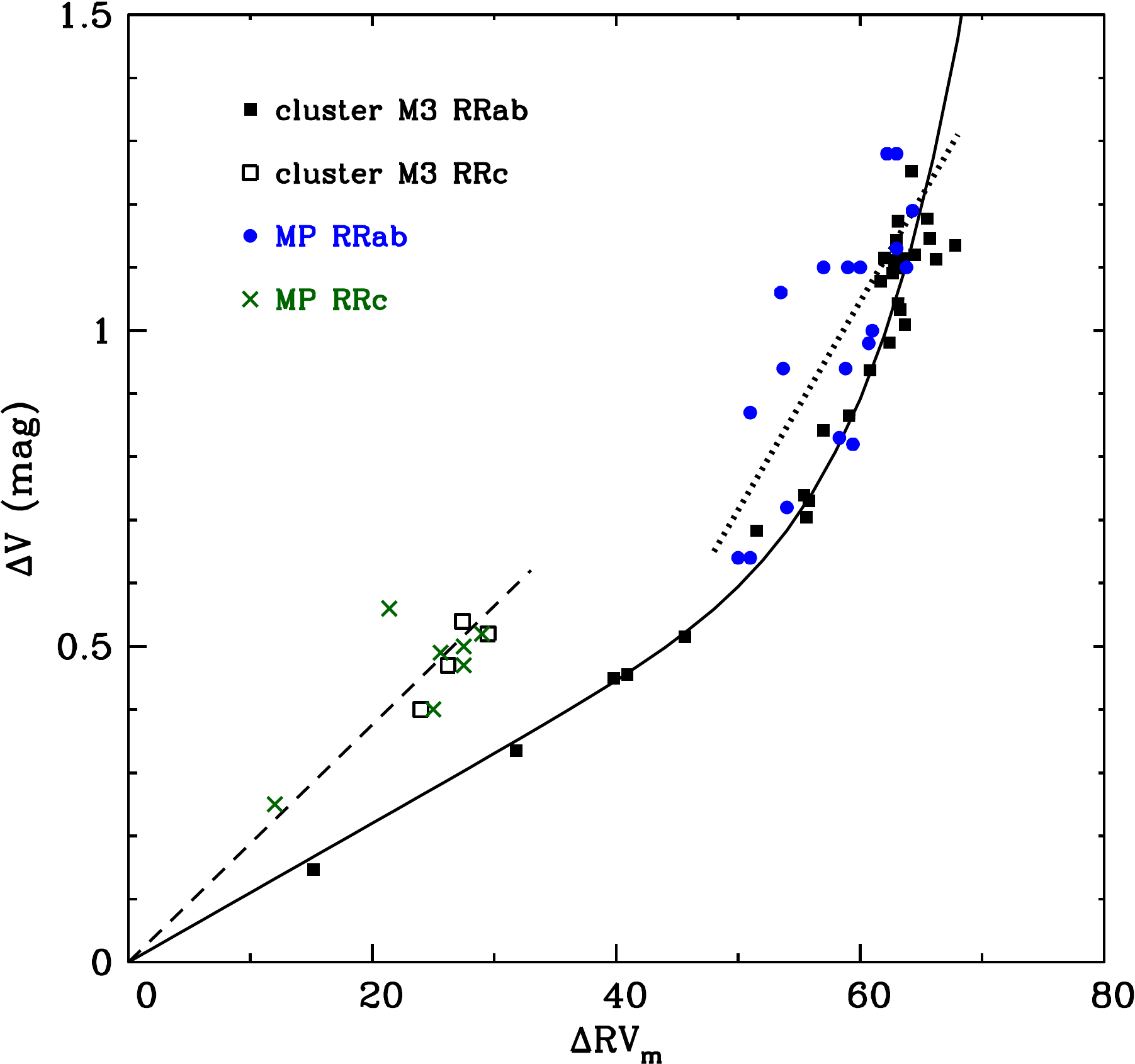}
\caption{\label{f8}
\footnotesize
   Visual light amplitude versus $RV$ amplitude.  
   The blue solid circles (MP RRab stars) and green crosses (MP RRc stars) 
   denote field RR Lyrae stars of this study. 
   Black points are RRab and RRc stars in globular cluster M3 
   ([Fe/H]~$\simeq$~$-$1.5) studied by \cite{jurcsik15,jurcsik17}.  
   The dashed regression line is for the RRc field stars, the dotted line
   is for the RRab field stars, and the solid line is for the M3 RRab stars.
   The field MR RRab stars are not shown in this figure.
}
\end{figure}

We combine the RRc photometric and $RV$ amplitudes of this study 
(Table~\ref{tab-stars}) with those for RRab stars in Table~4 of CSP17, 
and those for the variables in M3 recently investigated by 
\cite{jurcsik15,jurcsik17} to produce Figure~\ref{f8}.  
Regression lines through the origin 
($\Delta V$ = 0.019$\Delta RV$) for RRc stars,
linear 
($\Delta V$ = 0.033$\Delta RV$~$+$~0.934) for the RRab stars.
and non-linear power law
($\Delta V$ = 0.011$\Delta RV$~+~2.300$\times 10^{-17}\Delta RV^9$)
for the M3 observations of \citeauthor{jurcsik17} provide an 
observational picture of the different ways that fundamental and 
first overtone pulsations grow/vanish near the origin.

\begin{figure}
\epsscale{1.00}
\plotone{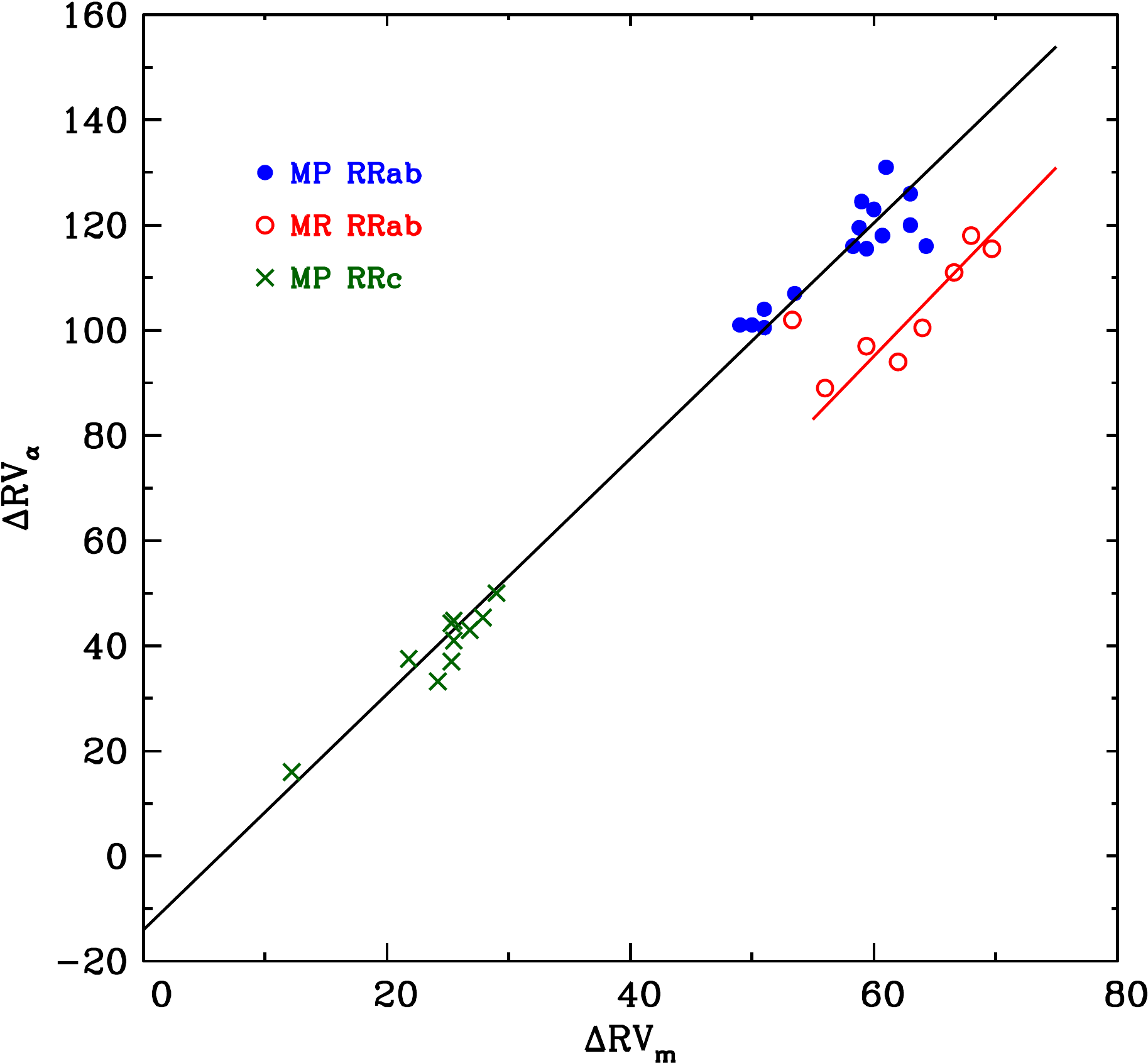}
\caption{\label{f9}
\footnotesize
   H$\alpha$ $RV$ amplitude $\Delta RV_\alpha$ plotted versus 
   metallic radial velocity amplitude $\Delta RV_m$.
   Point types and colors for the MP RRab and MP RRc stars are in 
   Figure~\ref{f8}; the red open circles are for MR RRab stars.
   The lines represent linear regression fits to the MR RRab data (red), 
   and the the combined MP RRab and MP RRc data.
}
\end{figure}

Similarly, we correlate the RRc metal and H$\alpha$
$RV$ amplitude data, $\Delta RV_m$ and $\Delta RV_\alpha$, for 
RRc and RRab stars to produce Figure~\ref{f9}.
The slopes of the MP and MR (metal-rich) regressions are similar, as in the 
preceding figure. 
We discuss these correlations further in \S\ref{metalhalpha}.

%%%%%%%%%%%%%%%%%%%%%%%%%%%%%%%%%%%%%%%%%%%%%%%%%%%%%%%%%%%%%%%%%%%%%%%%%%
\subsection{Radial Velocity Templates}\label{templates}
%%%%%%%%%%%%%%%%%%%%%%%%%%%%%%%%%%%%%%%%%%%%%%%%%%%%%%%%%%%%%%%%%%%%%%%%%%

Low-resolution spectroscopic surveys 
provide important data
about the density and kinematic structure of our Galaxy and its satellites.  
Single observations of RR~Lyrae stars with low-to-moderate spectral 
resolution can provide radial velocity accuracy ($\pm$5 to 10~\kmsec)
that is adequate for these purposes.  
However, the sizable $RV$ amplitudes of RR Lyrae stars during their pulsation 
cycles require special treatment (see \citealt{kollmeier13} for detailed 
discussion).  
If accurate photometric elements are available one may infer the 
center-of-mass velocity ($RV_{CoM}$) from the photometric phase of the 
observation by use of a velocity template.  
The template provides a correction that converts an observed $RV$ to $RV_{CoM}$.
Such templates have been provided for the metallic line $RV$s of RRab stars by 
\cite{liu91}.   
More recently \cite{sesar12} provided templates for RRab stars based on Balmer 
lines for use with low resolution spectra.  

The template correction procedure requires a photometric ephemeris, and it 
assumes a universal phase relation between times of light maximum and 
$RV$ minimum; for RRab stars the relationship is 
coincident at the $\pm$0.01$P$ level.  
As noted in \S\ref{radial} we cannot verify this coincidence for 
our RRc sample.  
We proceed with construction of templates in the expectation that a 
universal phase relation between light and velocity variations of the 
RRc stars will be forthcoming.  
Absent such a universal relation, there is no use for templates.

Omitting outlier AS143322, the remaining eight stable RRc stars in 
our sample comprise a homogeneous group with respect to visual light amplitude 
as discussed in \S\ref{radial}, so we thought it sufficient to create 
a single, scalable template for the metal lines of metal-poor RRc stars.   
In view of the significantly different behavior of H$\alpha$ radial
velocity we created a separate template for H$\alpha$.  
For the metal template we selected the seven stars with reasonably complete 
$RV$ phase coverage: AS023706, AS094541, AS123811, AS132225, AS132448, 
AS190212, and AS211933
(see Table~\ref{tab-stars} and Figures~\ref{f1} and \ref{f2}).  
We omitted AS143322, whose pulsation amplitudes are much smaller than all 
others in our sample.
For the H$\alpha$ template we also removed AS190212 because of the strange 
behavior of H$\alpha$ during rising light, and we added AS014500.

\begin{figure}
\epsscale{0.90}
\plotone{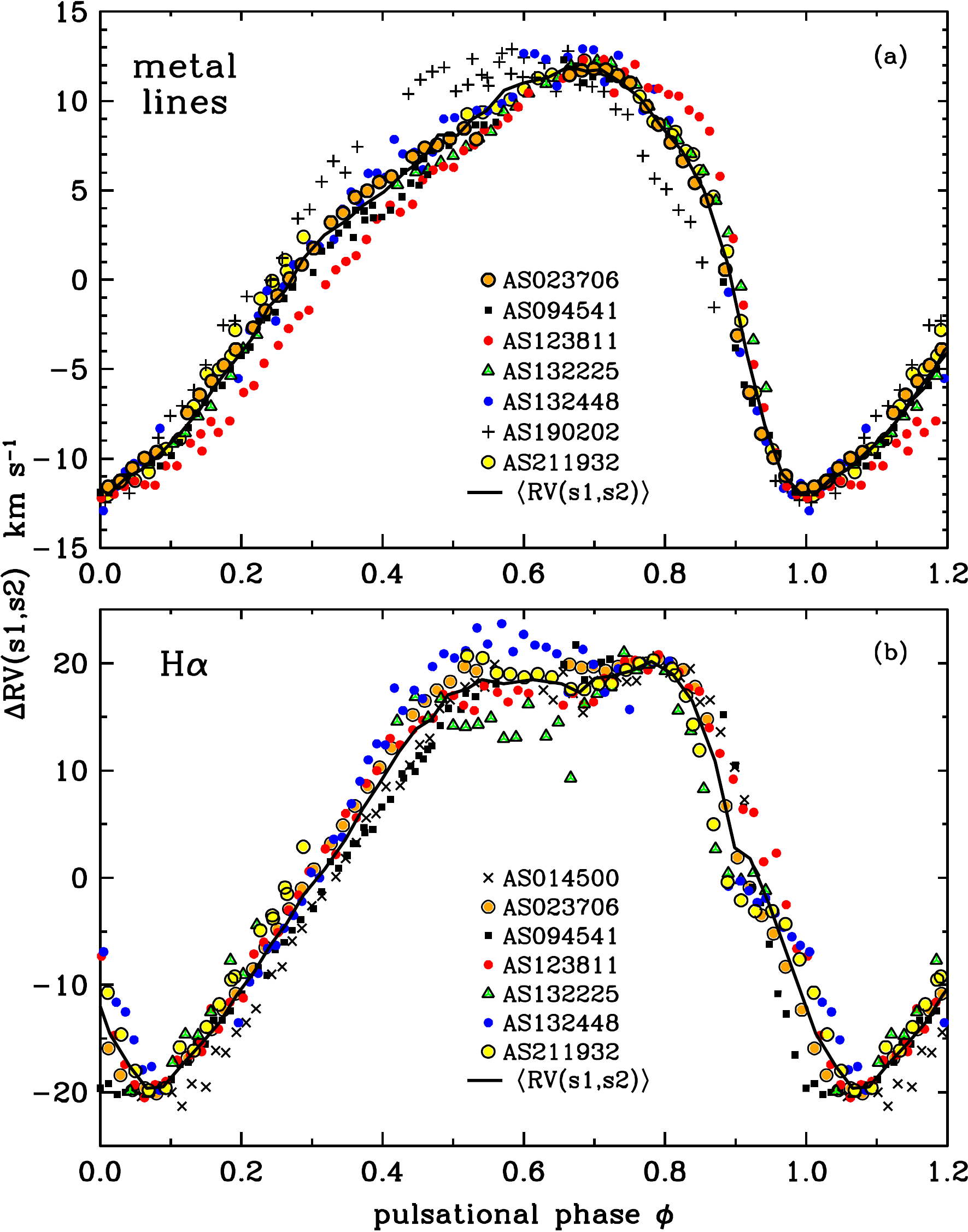}
\caption{\label{f10}
\footnotesize
   Stretched and shifted $RV$ variations for selected program stars 
   as functions of their pulsational phases.
   Individual measurements for each star are coded by symbols in the 
   legends for metal lines (panel a) and H$\alpha$ (panel b).  
   The solid black curves represent average values calculated as described 
   in the text.
   See the text for discussion of the stretching and shifting procedures.
}
\end{figure}

We superposed the metal $RV$ data for these stars, stretching/compressing 
measured $RV$ variations to two standard amplitudes, $\Delta RV$~=~24~\kmsec\
and 30~\kmsec, by use of a scale factor s1, which we centered on 
$RV$~=~0.0~\kmsec\ by use of a shift s2.  
The values of s1 and s2 that we used are given in Table~\ref{tab-shift}.
$RV$(s1,s2) denotes velocities created in this manner.  
The $RV$(s1,s2) data for $\Delta RV$are displayed in panel (a) of 
Figure~\ref{f10}, where various colored symbols are used to identify 
the contributions of each star.  
We define an average metal velocity variation for the RRc stars as follows.  
All the data values in Figure~\ref{f10} were sorted in order of 
increasing phase.  
Then average values of phase and $RV$(s1,s2) were calculated in bins of 10 
successive values.  
The average phase interval between successive average values 
$\langle RV$(s1,s2)$\rangle$ produced by this procedure was 
$\Delta\phi$~=~0.026~$\pm$~0.003.  
These are displayed by the continuous black curves in the figure.

It is evident from panel (a) of Figure~\ref{f10} 
that individual stars depart systematically from average metal line
$RV$ behavior: the trend for AS190212,
the most metal-poor star in our sample, is a conspicuous example.  
Departures of its $RV$(s1,s2) values from the mean curve as large as 5~\kmsec\
occur during about one third of its pulsation cycle.  
For most of the sample the deviations are much smaller.  
In view of the characteristic magnitudes of space motions and dispersions 
of Galactic RR~Lyrae stars ($\sim$100~\kmsec), we believe our 
mean relations provide a satisfactory way to convert individual $RV$
observations to $RV_{CoM}$ values for many purposes.

An identical procedure was used to create the diagram for H$\alpha$ in 
the panel (b) of Figure~\ref{f10}.
This diagram serves to call attention to the definition of the H$\alpha$ 
$RV$ amplitude, which we calculate as the difference between the $RV$
minimum that occurs near phase $\phi$~$\sim$~0.1 and the local velocity 
maximum near $\phi$~$\sim$~0.8.  
Data superposed in this manner produce a large range in a second local 
$RV$ maximum that occurs near $\phi$~$\sim$~0.5.  
We only note here that this secondary maximum has a morphological counterpart 
in the $RV$ curves of RRab stars near phase $\phi$~$\sim$~0.7 (CSP17). 
For some stars this maximum is larger (\eg, AS132448 and AS211933) than 
the one near $\phi$~$\sim$~0.8.

\begin{figure}
\epsscale{0.90}
\plotone{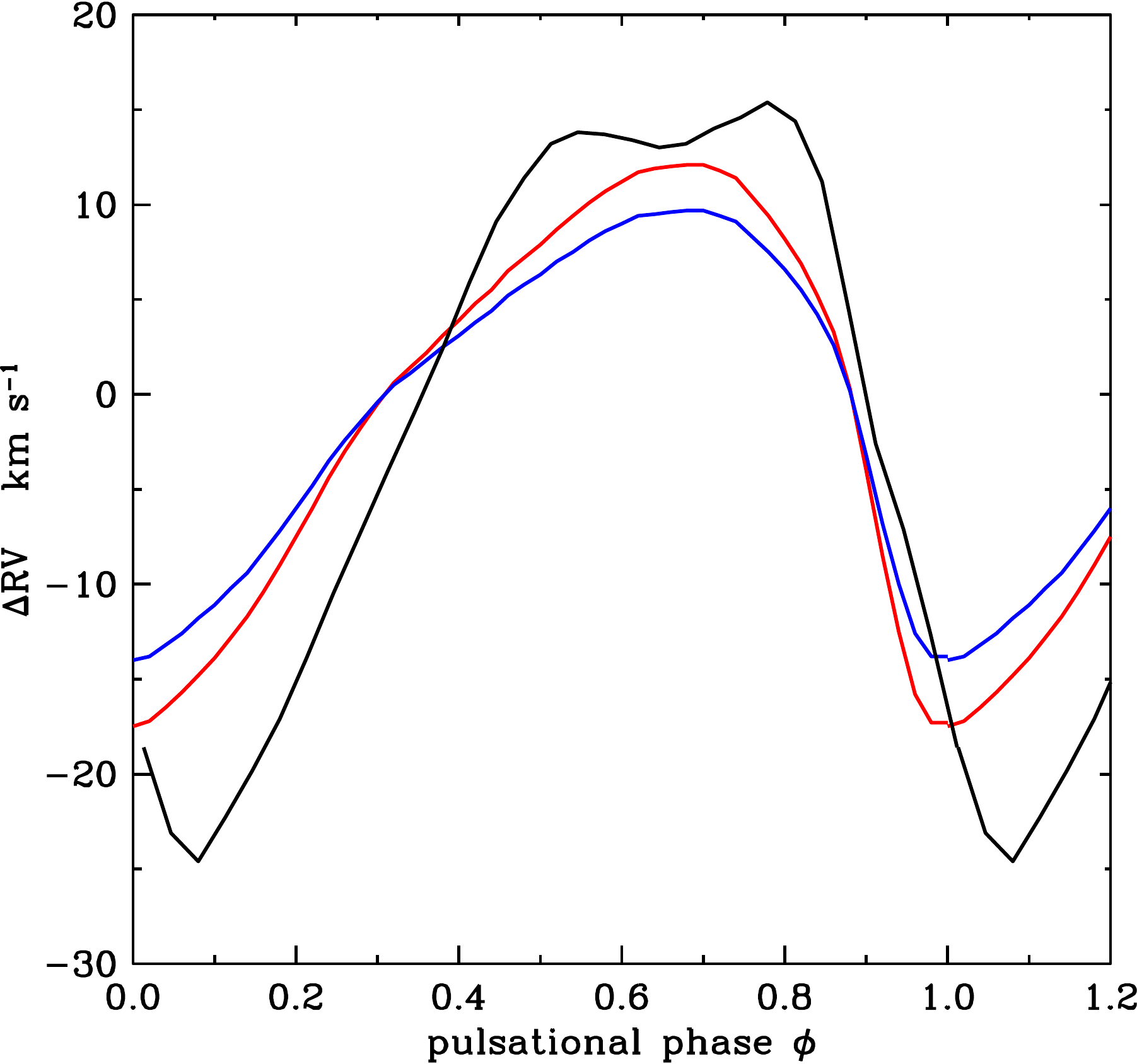}
\caption{\label{f11}
\footnotesize
   Mean $RV$ variations for metal lines (red and blue curves)
   and H$\alpha$ (black curve) derived from the data of Figure~\ref{f10}.
}
\end{figure}

We converted the mean $\langle RV$(s1,s2)$\rangle$ curves to templates by 
calculating the time-averaged values of these curves and shifting the zero 
points of the velocity scales to produce template velocity corrections.  
The results of these calculations are presented in Table~\ref{tab-curvemean}
and displayed in Figure~\ref{f11}.
In this table we tabulate $\Delta RV$~=~$RV(\phi$)~$-$~$RV_{CoM}$, the 
correction to be subtracted from a radial velocity observed at phase $\phi$.  
The tabulated corrections can be scaled to any amplitude. 
The time-averaged $RV$ of any template scaled in this manner remains 0.0~\kmsec. 
In Figure~\ref{f11} we show templates for metal line $RV$ 
amplitudes 24~\kmsec\ (blue line) and 30~\kmsec\ (red line).
These curves bracket the range of $\Delta RV$ possibilities in our sample. 
We also draw a black line with $RV$ amplitude 40~\kmsec\ for the typical 
variation of H$\alpha$ lines.

%%%%%%%%%%%%%%%%%%%%%%%%%%%%%%%%%%%%%%%%%%%%%%%%%%%%%%%%%%%%%%%%%%%%%%%%%%
\section{H$\alpha$ AND METALLIC-LINE VELOCITY VARIATIONS}\label{metalhalpha}
%%%%%%%%%%%%%%%%%%%%%%%%%%%%%%%%%%%%%%%%%%%%%%%%%%%%%%%%%%%%%%%%%%%%%%%%%%

\subsection{Photospheric Dynamics}

We used the metallic radial velocity data for the twelve stable RRc stars 
displayed in Figures~\ref{f1} and \ref{f2} to calculate pulsation 
velocities, center-of-mass ($\gamma$) velocities, radius variations, and 
primary accelerations (those that occur during primary light rise; see 
CSP17).

From the heliocentric radial velocity $RV(\phi)$, we calculate the 
pulsation velocity dR/dt($\phi$) in the stellar rest frame using:
d$R$/d$t(\phi$) = $-p(RV(\phi) - RV_\gamma$), where $p$ is the value of the 
correction factor for geometrical projection and limb darkening and 
$RV_\gamma$ is the center-of-mass velocity. 
Because the metallic absorption lines are formed near the photosphere, the 
pulsation velocity d$R$/d$t(\phi)$ is close to the photosphere motion 
and $RV_\gamma$ is assumed to be equal to the so-called $\gamma$-velocity, 
i.e., the average value of the heliocentric radial velocity curve over 
one pulsation period. 
The use of this approximation is discussed by \cite{bono94}. 
Here, we use $p$~=~1.36 from \cite{burki82} supported by recent studies 
\citep{neilson12,ngeo12}.

The pulsational velocities derived from metal lines are 
very nearly those of the photosphere: the metal line forming region at 
$\tau_{5000\AA}$~=~0.15 lies only $\sim$4~km above the photosphere of a 
\cite{kurucz11} model with the characteristic parameters of an RRc 
atmosphere (\teff~=~7250~K, \logg~=~2.5).

Radius and acceleration curves were computed from the integrals and 
derivatives of the pulsational velocity curves, respectively.
During pulsation cycles (0~$<$~$\phi$~$<$~1) gaps in the data were 
interpolated linearly. 
In order to minimize the noise, the raw radial velocity curves were 
convoluted with a sliding window of $\Delta$$\phi$~=~0.1 width.
Table~\ref{tab-rv} contains a summary of these photospheric motion parameters.

\begin{figure}
\epsscale{0.90}
\plotone{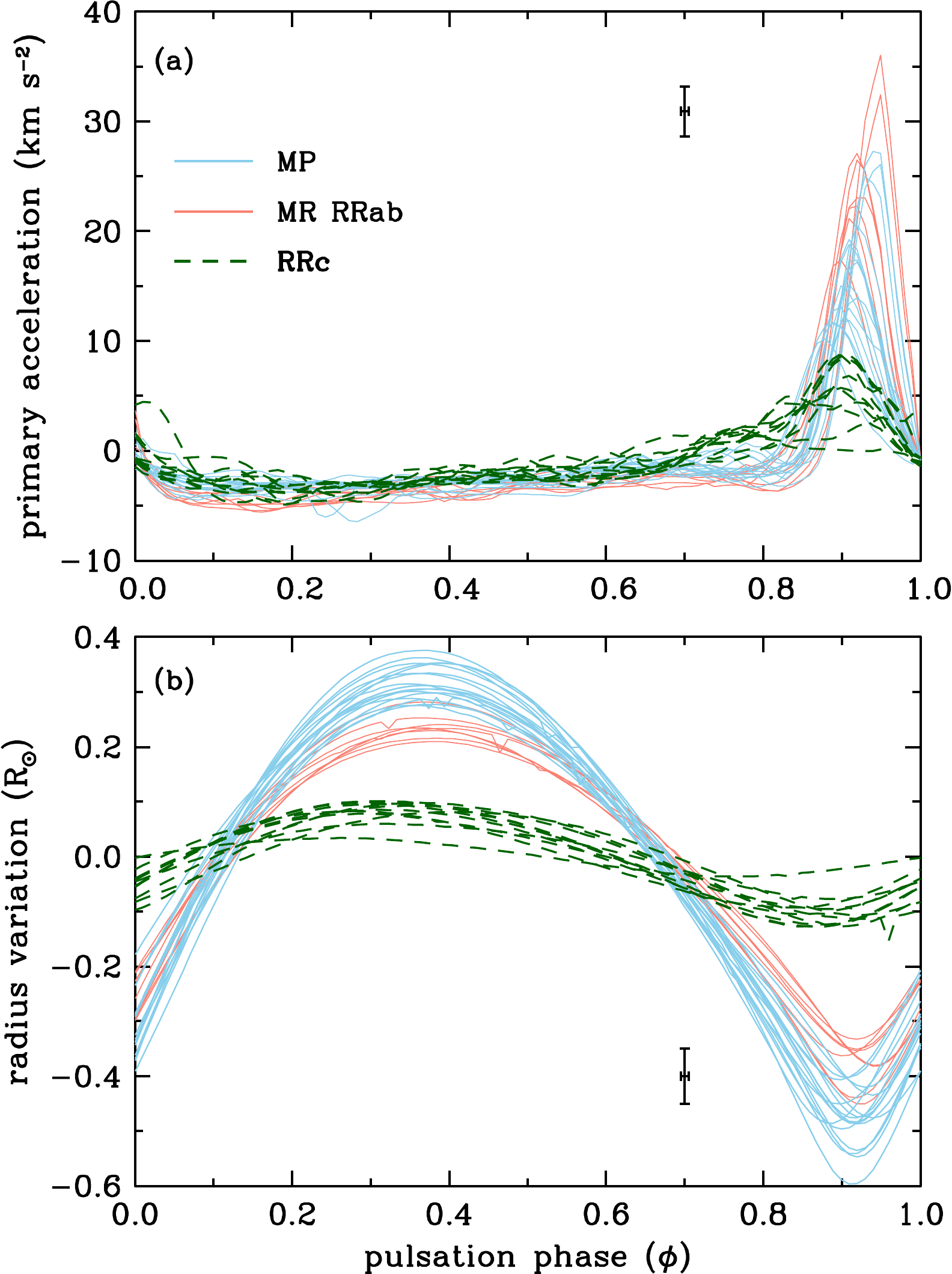}
\caption{\label{f12}
\footnotesize
   Dynamical acceleration and radius variation curves for individual RRc stars 
   (green dashed lines), metal-rich RRab stars (MR; light red lines) and 
   metal-poor RRab stars (MP; light blue lines).
}
\end{figure}

The acceleration and the radius variation curves for individual RRc stars are 
shown in Figure~\ref{f12}.
There is substantial star-to-star variation in acceleration maxima seen
in panel (a), with AS211933 reaching $\sim$~1.0~km~s$^{-2}$ while 
AS143322 exhibits very little variation in its acceleration throughout 
its pulsational phases.
In addition to the RRc curves we have added those for RRab stars from 
CSP17; see their Figure~11.
Maxima of the RRc acceleration curves are about half those 
of metal-poor RRab and about one third of those of metal-rich RRab.  
The peak acceleration (the dynamical gravity) occurs near phase 
$\phi$~=~0.90~$\pm$~0.05, during the phase of minimum radius displayed 
in panel (b) of Figure~\ref{f12}.

The data of Figure~\ref{f12} show that the radius variations 
and primary accelerations of all RRc stars are markedly smaller than 
those of metal-rich (MR) and metal-poor (MP) RRab stars.
In particular, the amplitudes of radius variations and the dynamical 
gravities of MP RRc stars are four times smaller than those of MP RRab stars.  
The time-average radii of RRc stars and those of MP and MR RRab 
stars occur at the same phases, $\phi$~=~0.65~$\pm$~0.05 during infall and 
$\phi$~=~0.12~$\pm$~0.05 during outflow.
All RRc stars of this study, except AS143322 which exhibits the lowest 
dynamical gravity, possess more or less prominent H$\alpha$ radial velocity 
maxima or at least stationary values near phase $\phi$~= 0.52~$\pm$~0.05
(Figures \ref{f10} and \ref{f11}).
This phenomenon, not seen in the metallic velocities, is restricted to the 
high atmosphere layers which produce Doppler cores of H$\alpha$.

\subsection{Shock Waves}\label{shocks}

The primary accelerations are due to the behaviors of radiative opacities, 
the $\kappa$ and $\gamma$ mechanisms, that operate in the He$^{+}$ and 
He$^{++}$ ionization zones that drive the pulsational instability.
These processes create compression waves and under certain conditions shock 
waves, $Sh_{H+He}$, that move outward through mass shells.
We have not detected any doubling of $H\alpha$ absorption lines or any 
H$\alpha$ emission during the primary accelerations of our RRc stars: 
shocks are weak or absent in the shallow RRc atmospheres. 
Failure to detect broadening of H$\alpha$ near phases of peak acceleration 
in our spectra sets an upper limit of approximately 10~\kmsec\ on any shock 
discontinuities that might be present.
We suppose instead that the prominent secondary radial velocity maxima or 
elbows near phase $\phi$~= 0.52~$\pm$~0.05 in Figures~\ref{f10} and 
\ref{f11}, are caused by compression waves.
These waves are qualitatively similar to $Sh_{PM3}$ in RRab stars described 
in \cite{chadid14}, the cooling shock \citep{hill72} that 
produces compression heating during infall. 
Such shocks are significant in extended atmospheres, leading to the 
prominent secondary maxima seen at H$\alpha$ $RV$ curves of metal-poor 
RRab stars.

\subsection{Comparison of RRc Atmospheric Structures with those of RRab Stars}

The linear regression of H$\alpha$ $RV$ amplitudes with those of metal 
lines in Figure~\ref{f9} is highly determinate ($R^2$~=~0.98), it 
appears to be strictly linear in the interval 
10~$<$~$\Delta RV_m$~$<$~65~\kmsec, and it has a 
``non-physical y-intercept'' of $-$14.5~\kmsec.  
The actual regression must become non-linear near the origin in a manner 
that permits both velocity amplitudes to approach zero together.  
This is a theoretical issue beyond the scope of our report.

Panel (a) of Figure~\ref{f12} illustrates the small dynamical 
gravities of RRc overtone pulsators.  
As noted in \S\ref{shocks} RRc stars may well be shock-free.
The data panel (a) of this figure suggests a dynamical gravity division 
between the RRc and RRab stars near $d^2R/dt^2$~=~15~km~s$^{-2}$.
Photospheric radius variations of RRc are similarly smaller than those of 
RRab stars (panel (b) of Figure 12).   
The RRc radius variation average is approximately 
$R$~$\simeq$~0.19~$R_\odot$.

\begin{figure}
\epsscale{0.75}
\plotone{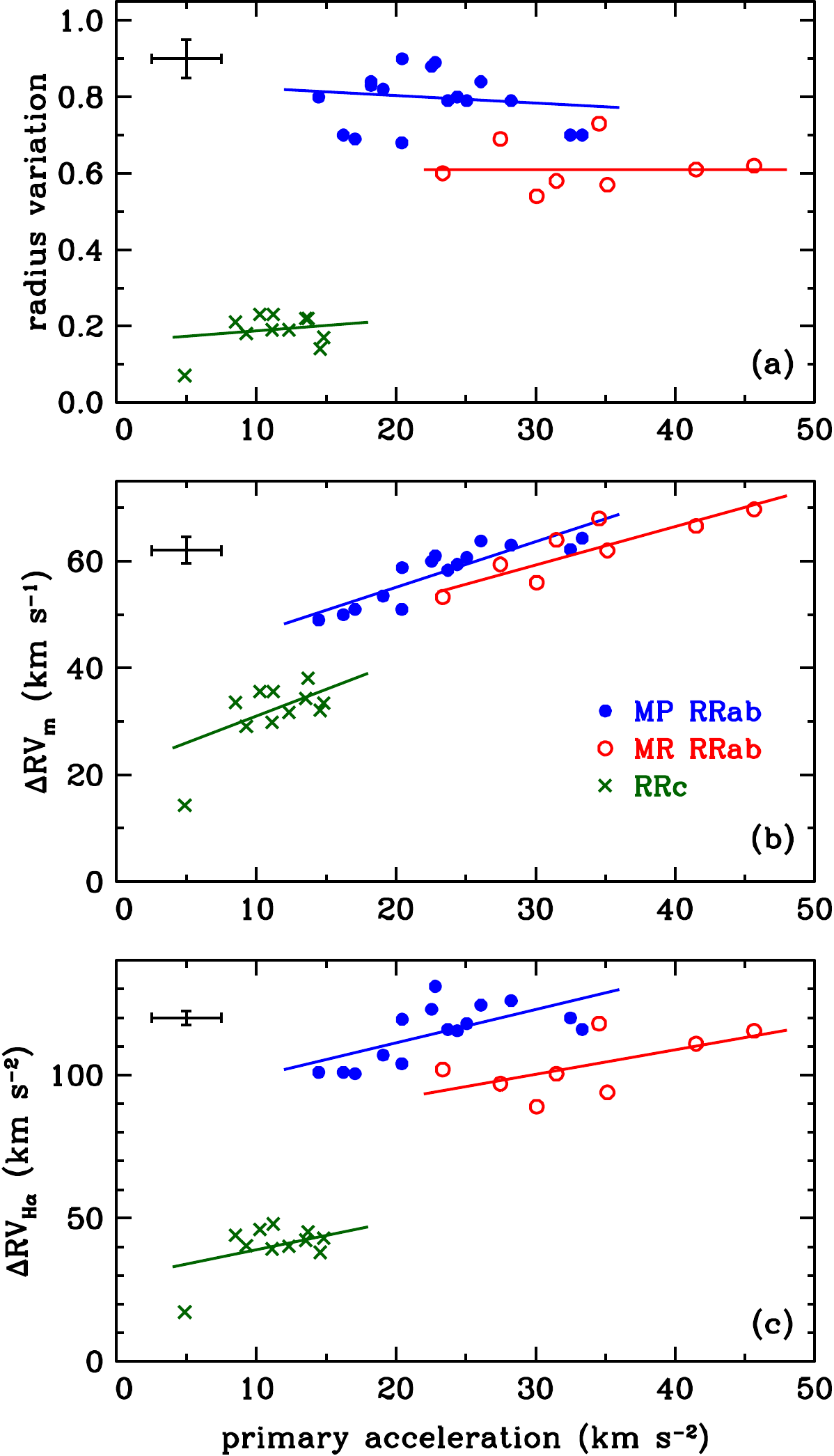}
\caption{\label{f13}
\footnotesize
   Radius variation, metal $RV$ amplitude  and H$\alpha$ $RV$ amplitude 
   $versus$ primary acceleration for RRc (green), metal-poor (blue) and 
   metal-rich (red) RRab stars.}
\end{figure}

Figure~\ref{f13} deserves some comment.  
Dynamical gravity, used as an independent variable, produces simple linear 
regressions for the radius variation, as well as the H$\alpha$ and 
metal line $RV$ amplitudes. 
We include the RRab data in CSP17 in all three panels of Figure~\ref{f13}
to show that the RRab and RRc stars lie on distinctly different regressions, 
i.e., motions of the upper-atmospheric and near-photospheric regions of 
RRc and RRab stars are distinctly different. 
Such differences in the atmospheric dynamics and structure are induced by 
large differences in the mechanical energy of the ballistic motion and 
the strengths of shock waves in their atmosphere. 
The RRc stars are trans-sonic regime stars (as characterized by CSP17);  
their pulsation excitation processes create atmospheric dynamics and 
structure that differ from those of the RRab stars.

%%%%%%%%%%%%%%%%%%%%%%%%%%%%%%%%%%%%%%%%%%%%%%%%%%%%%%%%%%%%%%%%%%%%%%%%%%
\section{TEST OF THE BLAZHKO ROTATOR HYPOTHESIS}\label{blazhko}
%%%%%%%%%%%%%%%%%%%%%%%%%%%%%%%%%%%%%%%%%%%%%%%%%%%%%%%%%%%%%%%%%%%%%%%%%%

Observed spectral line broadening is a convolution of axial rotation,
macroturbulence, microturbulence, thermal effects, and spectrograph slit
functions.
The last three of these causes can be estimated and subtracted from the total,
leaving a combination of macroturbulent and rotational line profiles.
\cite{preston13} describe in detail these calculations, calling the result
$V_{macrot}$. 
It is difficult to estimate the macroturbulence and rotation contributions
to $V_{macrot}$ unless one or the other of them dominates the broadening.
Here we attempt to detect rotational signatures in the line profiles of
Blazhko RRc stars.

The majority of Blazhko periods exceed 12 days, and in these cases any
line broadening produced by rotation would be unobservably small. 
To make our test we selected seven ASAS RRc stars for which the 
\cite{szczygiel07} Blazhko periods ($<$12~days) would produce axial 
equatorial rotational velocities $>$20~\kmsec, a value that we could measure 
from line profiles produced by the duPont echelle. 
To estimate these hypothetical axial rotations we adopted a mean radius 
$R_{RRc}/R_\odot$~=~4.56$\pm$0.42, which is the average of values obtained 
for two field RRc stars by use of the Baade-Wesselink method \citep{liu90} 
and values obtained from Fourier decomposition of photometry of two RRc stars 
in the globular cluster M2 \citep{lazaro06}.  
Then, from $2\pi R$~=~$V_{rot}P$ we obtain 
$V_{rot}$ (\kmsec)~=~50.6($R_{RRc}/R_\odot$)/$P$(d), which yields the $V_{rot}$ 
values in column~3 of Table~\ref{tab-pblvrot}.

Then following the procedures of \cite{preston13} we have computed 
$V_{macrot}$ values from the observed line profiles for the seven chosen RRc
stars, listing them in column~4 of Table~\ref{tab-pblvrot}.
These are upper limits to the rotational broadening: 
$V_{rot}$sin($i$)~$<$~$V_{macrot}$.
Thus, upper limits to sin($i$) can be computed (column~5 of 
Table~\ref{tab-pblvrot}), translating to upper limits on axial 
inclinations $i$ (column~6).
This distribution of maximum inclinations is improbable: if Blazhko periods 
were truly periods of axial rotation, we would have to be viewing Blazhko 
stars preferentially pole-on.  
We regard the improbability of this circumstance as falsification of the 
rotator hypothesis.  
We defer further discussion of this result to our next paper 
\citep{preston17}.

\acknowledgments

We thank all the Las Campanas Observatory support personnel for their help 
during the course of our endeavor.
We offer our special regards to several duPont telescope operators 
for their efforts in assisting with the observations required to produce 
this paper.  
We appreciate the help on IRAF spectral reductions given by Ian Thompson,
and discussion of warm variable stars with Michel Breger.
We give special thanks to Johanna Jurcsik for her careful reading of our 
initial manuscript, which resulted in several substantial improvements to 
the paper.
Finally, we are most grateful to Stephen Shectman for inventing the duPont 
echelle spectrograph thirty-some years ago.
This work has been supported in part by NSF grants AST-1211585 and 
AST1616040 to C.S, and M.A. acknowledges the Mobility+III fellowship 
from the Polish Ministry of Science and Higher Education. 

\facility{Carnegie Institution for Science (CIS) 2.5m (100 inch) Irenee 
Du Pont Telescope at Las Campanas Observatory (LCO)}

\software{ATLAS (Kurucz 2011), MOOG (Sneden 1973), 
          pyEW (https://github.com/madamow),
          pyMOOGi (https://github.com/madamow)}

%%%%%%%%%%%%%%%%%%%%%%%%%%%%%%%%%%%%%%%%%%%%%%%%%%%%%%%%%%%%%%%%%%%%%%%%%%
%    FIGURES
%%%%%%%%%%%%%%%%%%%%%%%%%%%%%%%%%%%%%%%%%%%%%%%%%%%%%%%%%%%%%%%%%%%%%%%%%%

\clearpage

\clearpage

\clearpage

\clearpage

%%%%%%%%%%%%%%%%%%%%%%%%%%%%%%%%%%%%%%%%%%%%%%%%%%%%%%%%%%%%%%%%%%%%%%%%%%   
%     TABLES
%%%%%%%%%%%%%%%%%%%%%%%%%%%%%%%%%%%%%%%%%%%%%%%%%%%%%%%%%%%%%%%%%%%%%%%%%%
 
\clearpage
\begin{center}
\begin{deluxetable}{cccccccc}
\tabletypesize{\footnotesize}
\tablewidth{0pt}
\tablecaption{Program Star Data\label{tab-stars}}
\tablecolumns{8}
\tablehead{
\colhead{ASAS Name\tablenotemark{a}}     &
\colhead{Other Name}                     &
\colhead{[Fe/H]\tablenotemark{b}}        &
\colhead{$P$\tablenotemark{c}}           &
\colhead{$HJD_0$\tablenotemark{c}}       &
\colhead{$V_{min}$}                      &
\colhead{$V_{amp}$}                      &
\colhead{$RV_{amp}$\tablenotemark{d}}    \\
\colhead{}                               &
\colhead{}                               &
\colhead{}                               &
\colhead{ASAS}                           &
\colhead{this study}                     &
\colhead{ASAS}                           &
\colhead{ASAS}                           &
\colhead{this study}
}
\startdata
\multicolumn{8}{c}{Stable RRc Stars} \\
014500-3003.6  &     SC Sci     & $-$2.28 &   0.377380 &   1869.075 &  11.14 &  0.52 &  \nodata  \\
023706-4257.8  &     CS Eri     & $-$1.88 &   0.311326 &   1868.717 &   8.84 &  0.50 &    29.0   \\
094541-0644.0  &     RU Sex     & $-$2.10 &   0.350225 &   1869.604 &  10.60 &  0.44 &    27.5   \\
095328+0203.5  &     T Sex      & $-$1.76 &   0.324697 &   1869.758 &   9.85 &  0.43 &  \nodata  \\
101332-0702.3  &                & $-$1.73 &   0.313619 &   1869.690 &  10.78 &  0.40 &  \nodata  \\
123811-1500.0  &     Y Crv      & $-$1.39 &   0.329045 &   1884.258 &  11.42 &  0.44 &    25.6   \\
132225-2042.3  &   BD-19 3673   & $-$0.96 &   0.235934 &   1888.536 &  10.72 &  0.40 &    25.0   \\
132448-0658.8  &     AU Vir     & $-$2.04 &   0.323237 &   1900.042 &  11.43 &  0.51 &    27.5   \\
143322-0418.2  &   BD-03 3640   & $-$1.48 &   0.249632 &   1907.348 &  10.61 &  0.23 &    12.0   \\
190212-4639.2  &     MT Tel     & $-$2.58 &   0.316690 &   1954.795 &   8.75 &  0.55 &    21.4   \\
203145-2158.7  &                & $-$1.17 &   0.310712 &   1873.190 &  11.27 &  0.38 &  \nodata  \\
211933-1507.0  &     YZ Cap     & $-$1.50 &   0.273460 &   1873.307 &  11.11 &  0.50 &    25.6   \\
\multicolumn{8}{c}{Blazhko RRc Stars} \\
081933-2358.2  &    V701 Pup    & $-$2.50 &   0.285667 &   4900.180 &  10.42 &  0.28 &  \nodata  \\
085254-0300.3  &                & $-$1.50 &   0.266902 &   4900.400 &  12.46 &  0.47 &  \nodata  \\
090900-0410.4  &                & $-$1.98 &   0.303261 &   4900.200 &  10.69 &  0.41 &  \nodata  \\
110522-2641.0  &                & $-$1.60 &   0.294510 &   4900.150 &  11.66 &  0.39 &  \nodata  \\
162158+0244.5  &                & $-$1.83 &   0.323698 &   4900.435 &  12.62 &  0.40 &  \nodata  \\
200431-5352.3  &                & $-$2.66 &   0.300240 &   4900.470 &  11.01 &  0.28 &  \nodata  \\
230659-4354.6  &     BO Gru     & $-$1.24 &   0.281130 &   5014.925 &  12.79 &  0.30 &  \nodata  \\
\enddata

\tablenotetext{a}{All Sky Automated Survey \citep{pojmanski03}; in the text 
                  these names will be shortened to the first set of digits, 
                  \eg, AS014500-3003.6 will be referred to as AS014500}
\tablenotetext{b}{the mean of Table~\ref{tab-total} entries for neutral and
                  ionized Fe species}
\tablenotetext{c}{for $P$ the source is ASAS for the stable RRc stars,
                  and \cite{govea14} for the Blazhko stars; for $HJD0$
                  the source is this study for the stable stars, and
                  \cite{govea14} for the Blazhko stars}
\tablenotetext{d}{$RV$ amplitudes of the metal lines}
\end{deluxetable}
\end{center}

\begin{center}
\begin{deluxetable}{cccccc}
\tabletypesize{\footnotesize}\tablewidth{0pt}
\tablecaption{Radial Velocities\label{tab-rvobs}}
\tablecolumns{6}
\tablehead{
\colhead{Star}                    &
\colhead{file}                    &
\colhead{HJD\tablenotemark{a}}    &
\colhead{$\phi$}                  &
\colhead{RV(metals)}              &
\colhead{RV(H$\alpha$)}           \\
\colhead{}                        &
\colhead{}                        &
\colhead{d}                       &
\colhead{}                        &
\colhead{\kmsec}                  &
\colhead{\kmsec}
}
\startdata
AS014500 & ccd6630  & 6916.7317 & 0.059 & $-$29.1 & $-$43.6 \\
AS014500 & ccd6631  & 6916.7370 & 0.074 & $-$29.5 & $-$43.0 \\
AS014500 & ccd6632  & 6916.7424 & 0.088 & $-$27.9 & $-$42.4 \\
AS014500 & ccd6633  & 6916.7477 & 0.102 & $-$27.2 & $-$43.1 \\
AS014500 & ccd6634  & 6916.7531 & 0.116 & $-$26.7 & $-$44.6 \\
AS014500 & ccd6635  & 6916.7584 & 0.130 & $-$25.5 & $-$42.3 \\
AS014500 & ccd6637  & 6916.7659 & 0.150 & $-$25.9 & $-$42.6 \\
AS014500 & ccd6638  & 6916.7712 & 0.164 & $-$24.1 & $-$38.8 \\
AS014500 & ccd6639  & 6916.7765 & 0.178 & $-$23.2 & $-$39.0 \\
AS014500 & ccd6640  & 6916.7819 & 0.192 & $-$22.4 & $-$36.9 \\
\enddata

\tablenotetext{a}{+2450000 d}
(This table is available in its entirety in machine-readable form.)

\end{deluxetable}
\end{center}

\clearpage
\begin{center}
\begin{deluxetable}{crrrc}
\tabletypesize{\footnotesize}
\tablewidth{0pt}
\tablecaption{Phases of Radial Velocity and Photometric Minimas\label{tab-coin2}}
\tablecolumns{5}
\tablehead{
\colhead{Star}                                   &
\colhead{$\phi_{min}$}                           &
\colhead{$\phi_{min}$}                           &
\colhead{$\Delta\phi$\tablenotemark{a}}          &
\colhead{source}                                 \\
\colhead{}                                       &
\colhead{$RV$}                                   &
\colhead{$V_{mag}$}                              &
\colhead{}                                       &
\colhead{}                                       
}
\startdata
\multicolumn{5}{c}{RRab Stars}                           \\
\mbox{TW Her}    &     1.010 &     1.005 &     0.005 & 1 \\
\mbox{UU Vir}    &     1.030 &     1.000 &     0.030 & 1 \\
\mbox{V445 Oph}  &     1.000 &     1.000 &     0.000 & 2 \\
\mbox{SS Leo}    &     1.020 &     1.000 &     0.020 & 2 \\
\mbox{WY Ant}    &     0.990 &     1.000 &  $-$0.010 & 3 \\
\mbox{W Crt}     &     1.010 &     1.000 &     0.010 & 3 \\
\mbox{BB Pup}    &     1.025 &     1.000 &     0.025 & 3 \\
\mbox{SW And}    &     1.040 &     1.000 &     0.040 & 4 \\
\mbox{SW Dra}    &     1.025 &     1.000 &     0.025 & 4 \\
\mbox{SS For}    &     1.020 &     1.005 &     0.015 & 4 \\
\mbox{RV Phe}    &     1.010 &     1.005 &     0.005 & 4 \\
\mbox{V440 Sgr}  &     1.000 &     1.010 &  $-$0.010 & 4 \\
\mbox{SW And}    &     1.020 &     1.005 &     0.015 & 5 \\
\mbox{RR Cet}    &     1.015 &     0.985 &     0.030 & 5 \\
\mbox{SU Dra}    &     1.000 &     0.995 &     0.005 & 5 \\
\mbox{RX Eri}    &     1.000 &     1.000 &     0.000 & 5 \\
\mbox{TT Lyn}    &     1.000 &     1.000 &     0.000 & 5 \\
\mbox{AV Peg}    &     1.000 &     1.000 &     0.000 & 5 \\
\mbox{AR Per}    &     1.005 &     1.000 &     0.005 & 5 \\
\mbox{TU UMa}    &     1.020 &     0.995 &     0.025 & 5 \\
mean             &           &           &     0.012 &   \\
$\sigma$         &           &           &     0.014 &   \\
\multicolumn{5}{c}{RRc Stars}                           \\
\mbox{DH Peg}    &     1.045 &     0.995 &     0.050 & 6 \\
\mbox{T Sex}     &     1.000 &     0.990 &     0.010 & 5 \\
\mbox{v056}      &     0.000 &    -0.005 &     0.005 & 7 \\
\mbox{v086}      &     0.005 &     0.000 &     0.005 & 7 \\
\mbox{v097}      &  $-$0.010 &    -0.020 &     0.010 & 7 \\
\mbox{v107}      &     0.040 &     0.010 &     0.030 & 7 \\
mean             &           &           &     0.018 &   \\
$\sigma$         &           &           &     0.014 &   \\
\enddata

\tablenotetext{a}{$\phi_{min}$($RV$)~$-$~$\phi_{min}$($V_{mag}$)}
\tablerefs{1 - \cite{jones88b}; 2 - \cite{fernley90}; 
           3 - \cite{skillen93}; 4 - \cite{cacciari87}; 
           5 - \cite{liu89}; 6 - \cite{jones88a}:
           7 - \cite{jurcsik15,jurcsik17}}

\end{deluxetable}
\end{center}

\clearpage
\begin{center}
\begin{deluxetable}{cccccccccccccccc}
\tabletypesize{\footnotesize}
\tablewidth{0pt}
\tablecaption{Stellar Parameters from Phased-Averaged Spectra\label{tab-phase}}
\tablecolumns{16}
\tablehead{
\colhead{Star}                           &
\colhead{$\phi$}                         &
\colhead{num\tablenotemark{a}}           &
\colhead{\teff}                          &
\colhead{\logg}                          &
\colhead{[M/H]}                          &
\colhead{\vmicro}                        &
\colhead{[Fe/H]}                         &
\colhead{$\sigma$}                       &
\colhead{\#\tablenotemark{b}}            &   
\colhead{[Fe/H]}                         &
\colhead{$\sigma$}                       &
\colhead{\#}                             \\
\colhead{}                               &
\colhead{}                               &
\colhead{}                               &
\colhead{K}                              &
\colhead{}                               &
\colhead{}                               &
\colhead{\kmsec}                         &
\colhead{I}                              &
\colhead{I}                              &
\colhead{I}                              &
\colhead{II}                             &
\colhead{II}                             &
\colhead{II}    
}
\startdata
AS014500 &    0.055 &        6 &     7650 &     3.70 &    -2.00 &      3.0 &    -1.98 &     0.29 &       16 &    -1.94 &     0.22 &       11 \\
AS014500 &    0.146 &        6 &     7500 &     3.50 &    -2.00 &      3.0 &    -1.95 &     0.12 &       24 &    -1.95 &     0.26 &       14 \\
AS014500 &    0.247 &        7 &     7200 &     2.90 &    -2.00 &      2.5 &    -2.04 &     0.20 &       26 &    -2.01 &     0.22 &       19 \\ 
AS014500 &    0.353 &        7 &     6700 &     2.80 &    -2.00 &      2.5 &    -2.09 &     0.29 &       34 &    -2.05 &     0.24 &       22 \\ 
AS014500 &    0.451 &        6 &     6950 &     3.00 &    -2.00 &      3.2 &    -2.09 &     0.22 &       26 &    -2.12 &     0.26 &       17 \\
AS014500 &    0.566 &        5 &     6950 &     3.00 &    -2.00 &      2.6 &    -1.98 &     0.25 &       24 &    -2.00 &     0.26 &       15 \\       
AS014500 &    0.661 &        5 &     7050 &     3.20 &    -2.00 &      2.5 &    -2.02 &     0.21 &       22 &    -1.99 &     0.20 &       13 \\
AS014500 &    0.743 &        6 &     7650 &     3.50 &    -2.00 &      2.5 &    -1.99 &     0.18 &       15 &    -2.00 &     0.18 &       11 \\
AS014500 &    0.834 &        6 &     7650 &     4.00 &    -2.00 &      2.5 &    -2.01 &     0.18 &        8 &    -1.97 &     0.26 &        5 \\    
AS023706 &    0.049 &        5 &     7150 &     2.10 &     1.90 &      2.3 &    -1.86 &     0.17 &       48 &    -1.86 &     0.14 &       33 \\       
AS023706 &    0.161 &        5 &     7000 &     2.10 &    -1.90 &      2.0 &    -1.82 &     0.14 &       61 &    -1.80 &     0.12 &       35 \\
AS023706 &    0.255 &        5 &     6850 &     2.10 &    -1.90 &      2.3 &    -1.84 &     0.12 &       64 &    -1.82 &     0.13 &       38 \\ 
AS023706 &    0.355 &        6 &     6650 &     2.10 &    -1.90 &      2.5 &    -1.85 &     0.13 &       74 &    -1.85 &     0.14 &       39 \\
AS023706 &    0.461 &        5 &     6650 &     2.40 &    -1.90 &      2.4 &    -1.79 &     0.14 &       74 &    -1.79 &     0.15 &       35 \\
AS023706 &    0.589 &        5 &     6650 &     2.40 &    -1.90 &      2.4 &    -1.83 &     0.14 &       64 &    -1.82 &     0.15 &       33 \\
AS023706 &    0.747 &        6 &     7000 &     2.60 &    -1.80 &      2.6 &    -1.77 &     0.14 &       60 &    -1.79 &     0.12 &       35 \\
AS023706 &    0.857 &        6 &     7300 &     2.60 &     1.80 &      2.0 &    -1.87 &     0.14 &       35 &    -1.88 &     0.18 &       27 \\
AS023706 &    0.959 &        5 &     7300 &     2.60 &    -1.80 &      2.2 &    -1.89 &     0.16 &       39 &    -1.87 &     0.11 &       28 \\
\enddata

\tablenotetext{a}{num = the number of spectra co-added at this mean phase}
\tablenotetext{b}{\# = the number of lines contributing to an abundance mean}
(This table is available in its entirety in machine-readable form.)

\end{deluxetable}
\end{center}

\clearpage
\begin{center}
\begin{deluxetable}{ccccccccccccc}
\tabletypesize{\footnotesize}
\tablewidth{0pt}
\tablecaption{Mean Stellar Parameters: Phased-Averaged Spectra\label{tab-phasemean}}
\tablecolumns{13}
\tablehead{
\colhead{Star}                           &
\colhead{$\phi$}                         &
\colhead{num\tablenotemark{a}}           &
\colhead{\teff}                          &
\colhead{\logg}                          &
\colhead{[M/H]}                          &
\colhead{\vmicro}                        &
\colhead{[Fe/H]}                         &
\colhead{$\sigma$}                       &
\colhead{\#\tablenotemark{b}}            &
\colhead{[Fe/H]}                         &
\colhead{$\sigma$}                       &
\colhead{\#}                             \\
\colhead{}                               &
\colhead{}                               &
\colhead{}                               &
\colhead{K}                              &
\colhead{}                               &
\colhead{}                               &
\colhead{\kmsec}                         &
\colhead{I}                              &
\colhead{I}                              &
\colhead{I}                              &
\colhead{II}                             &
\colhead{II}                             &
\colhead{II}
}
\startdata
AS014500 &    0.44 &       6 &     7256 &     3.29 &    -2.00 &      2.7 &    -2.02 &     0.05 &       22 &    -2.00 &     0.05 &       14 \\ 
AS023706 &    0.50 &       5 &     6950 &     2.33 &    -1.04 &      2.3 &    -1.84 &     0.04 &       58 &    -1.83 &     0.03 &       34 \\      
AS081933 &    0.59 &       3 &     7400 &     3.70 &    -2.35 &      1.8 &    -2.50 &     0.03 &       12 &    -2.53 &     0.01 &        9 \\  
AS085254 &    0.61 &       2 &     7267 &     2.63 &    -1.53 &      2.9 &    -1.60 &     0.05 &       24 &    -1.61 &     0.07 &       20 \\   
AS090900 &    0.14 &       3 &     7300 &     2.50 &    -1.80 &      2.9 &    -1.93 &     0.05 &       18 &    -1.95 &     0.04 &       13 \\ 
AS094541 &    0.42 &       8 &     6963 &     2.60 &    -2.00 &      2.5 &    -1.98 &     0.04 &       27 &    -1.98 &     0.04 &       18 \\   
AS095328 &    0.35 &       8 &     6958 &     2.12 &    -1.48 &      2.3 &    -1.48 &     0.10 &       61 &    -1.49 &     0.11 &       34 \\ 
AS101332 &    0.39 &       9 &     6736 &     2.19 &    -1.50 &      3.1 &    -1.48 &     0.07 &       54 &    -1.47 &     0.08 &       31 \\    
AS110522 &    0.55 &       3 &     7175 &     2.88 &    -1.68 &      2.1 &    -1.73 &     0.16 &       23 &    -1.73 &     0.17 &       17 \\ 
AS123811 &    0.49 &       7 &     7360 &     2.90 &    -1.23 &      3.0 &    -1.21 &     0.09 &       34 &    -1.22 &     0.08 &       24 \\
AS132225 &    0.44 &       6 &     6979 &     2.27 &    -0.71 &      3.0 &    -0.96 &     0.04 &       56 &    -0.96 &     0.03 &       30 \\ 
AS132448 &    0.50 &       6 &     7050 &     2.50 &    -1.98 &      2.4 &    -1.92 &     0.24 &       15 &    -1.93 &     0.25 &        9 \\
AS143322 &    0.46 &       6 &     7311 &     2.77 &    -1.29 &      3.0 &    -1.26 &     0.12 &       99 &    -1.25 &     0.12 &       34 \\
AS162158 &    0.35 &       4 &     7408 &     3.48 &    -1.53 &      2.8 &    -1.56 &     0.11 &       13 &    -1.58 &     0.08 &       11 \\
AS190212 &    0.48 &       6 &     7270 &     3.23 &    -2.71 &      2.1 &    -2.70 &     0.16 &       20 &    -2.71 &     0.17 &       13 \\
AS200431 &    0.47 &       4 &     7486 &     3.37 &    -1.86 &      2.0 &    -2.42 &     0.19 &       10 &    -2.40 &     0.18 &        7 \\
AS203145 &    0.44 &       5 &     6663 &     2.20 &    -0.59 &      3.3 &    -1.09 &     0.06 &       66 &    -1.07 &     0.05 &       30 \\
AS211933 &    0.52 &       5 &     7044 &     2.15 &    -1.46 &      2.5 &    -1.49 &     0.05 &       66 &    -1.46 &     0.06 &       30 \\
AS230659 &    0.45 &       5 &     7550 &     2.93 &    -1.48 &      2.8 &    -1.58 &     0.07 &       15 &    -1.56 &     0.12 &       13 \\
\enddata

\tablenotetext{a}{num = the number of phase results averaged together}
\tablenotetext{b}{\# = the average number of lines from each phase that
contribute to an abundance mean}

\end{deluxetable}
\end{center}

\clearpage
\begin{center}
\begin{deluxetable}{ccccccccccccc}
\tabletypesize{\footnotesize}
\tablewidth{0pt}
\tablecaption{Mean Stellar Parameters: Total Spectra\label{tab-total}}
\tablecolumns{13}
\tablehead{
\colhead{Star}                           &
\colhead{$\phi$}                         &
\colhead{num\tablenotemark{a}}           &
\colhead{\teff}                          &
\colhead{\logg}                          &
\colhead{[M/H]}                          &
\colhead{\vmicro}                        &
\colhead{[Fe/H]}                         &
\colhead{$\sigma$}                       &
\colhead{\#\tablenotemark{b}}            &
\colhead{[Fe/H]}                         &
\colhead{$\sigma$}                       &
\colhead{\#}                             \\
\colhead{}                               &
\colhead{}                               &
\colhead{}                               &
\colhead{K}                              &
\colhead{}                               &
\colhead{}                               &
\colhead{\kmsec}                         &
\colhead{I}                              &
\colhead{I}                              &
\colhead{I}                              &
\colhead{II}                             &
\colhead{II}                             &
\colhead{II}
}
\startdata
AS014500  &    all   &       54 &     6950 &     2.50 &  $-$2.00 &     2.2 &  $-$2.30 &     0.19 &       34 &  $-$2.26 &     0.19 &       23 \\    
AS023706  &    all   &       48 &     6850 &     2.20 &  $-$1.90 &     2.1 &  $-$1.90 &     0.12 &       63 &  $-$1.87 &     0.13 &       35 \\
AS081933  &    all   &        6 &     7550 &     3.00 &  $-$2.50 &     1.0 &  $-$2.48 &     0.14 &       15 &  $-$2.52 &     0.15 &       12 \\
AS085254  &    all   &        6 &     7400 &     2.80 &  $-$1.50 &     2.2 &  $-$1.49 &     0.19 &       31 &  $-$1.51 &     0.22 &       26 \\ 
AS090900  &    all   &        6 &     7150 &     2.40 &  $-$1.90 &     2.5 &  $-$1.98 &     0.20 &       18 &  $-$1.97 &     0.15 &       17 \\
AS094541  &    all   &       67 &     6900 &     2.60 &  $-$2.00 &     2.0 &  $-$2.11 &     0.17 &       36 &  $-$2.10 &     0.15 &       20 \\    
AS095328  &    all   &       50 &     7000 &     2.30 &  $-$1.40 &     1.2 &  $-$1.76 &     0.14 &       57 &  $-$1.76 &     0.15 &       35 \\
AS101332  &    all   &       62 &     6600 &     2.10 &  $-$1.60 &     2.0 &  $-$1.74 &     0.18 &       62 &  $-$1.72 &     0.17 &       36 \\ 
AS110522  &    all   &       18 &     7500 &     3.20 &  $-$1.60 &     1.5 &  $-$1.61 &     0.20 &       32 &  $-$1.58 &     0.14 &       25 \\  
AS123811  &    all   &       65 &     7200 &     2.70 &  $-$1.40 &     2.5 &  $-$1.41 &     0.23 &       57 &  $-$1.37 &     0.19 &       35 \\
AS132225  &    all   &       39 &     7100 &     2.40 &  $-$1.00 &     2.0 &  $-$0.98 &     0.16 &       77 &  $-$0.95 &     0.16 &       39 \\  
AS132448  &    all   &       58 &     7100 &     2.70 &  $-$2.00 &     1.5 &  $-$2.04 &     0.20 &       27 &  $-$2.03 &     0.23 &       24 \\ 
AS143322  &    all   &       53 &     7100 &     2.30 &  $-$1.40 &     2.0 &  $-$1.47 &     0.13 &       65 &  $-$1.48 &     0.12 &       32 \\
AS162158  &    all   &       21 &     7600 &     3.20 &  $-$1.80 &     2.5 &  $-$1.83 &     0.19 &       22 &  $-$1.83 &     0.15 &       17 \\
AS190212  &    all   &       59 &     7500 &     3.70 &  $-$2.60 &     1.7 &  $-$2.59 &     0.05 &       20 &  $-$2.56 &     0.10 &       14 \\
AS200431  &    all   &       29 &     7600 &     3.20 &  $-$2.60 &     3.0 &  $-$2.67 &     0.14 &       14 &  $-$2.65 &     0.21 &       13 \\
AS203145  &    all   &       38 &     6600 &     1.90 &  $-$1.10 &     3.0 &  $-$1.18 &     0.18 &       77 &  $-$1.16 &     0.17 &       37 \\
AS211933  &    all   &       43 &     7200 &     2.20 &  $-$1.50 &     2.3 &  $-$1.51 &     0.13 &       58 &  $-$1.50 &     0.15 &       37 \\
AS230659  &    all   &       28 &     7750 &     3.10 &  $-$1.40 &     1.2 &  $-$1.26 &     0.25 &       33 &  $-$1.23 &     0.22 &       25 \\
\enddata

\tablenotetext{a}{num = the total number of spectra obtained for this star 
                  at all phases}
\tablenotetext{b}{\# = the number of lines contribute to an abundance}

\end{deluxetable}
\end{center}

\clearpage
\begin{center}
\begin{deluxetable}{ccccccccccccccccccc}
\tabletypesize{\footnotesize}
\tablewidth{0pt}
\tablecaption{Relative Abundances of $\alpha$ Elements\label{tab-alphas}}
\tablecolumns{19}
\tablehead{
\colhead{Star}                           &
\colhead{[Mg/Fe]}                        &
\colhead{$\sigma$\tablenotemark{a}}      &
\colhead{\#\tablenotemark{b}}            &
\colhead{[Si/Fe]}                        &
\colhead{$\sigma$}                       &
\colhead{\#}                             &
\colhead{[Si/Fe]}                        &
\colhead{$\sigma$}                       &
\colhead{\#}                             &
\colhead{[Ca/Fe]}                        &
\colhead{$\sigma$}                       &
\colhead{\#}                             &
\colhead{[Ti/Fe]}                        &
\colhead{$\sigma$}                       &
\colhead{\#}                             &
\colhead{[Ti/Fe]}                        &
\colhead{$\sigma$}                       &
\colhead{\#}                             \\
\colhead{}                               &
\colhead{I\tablenotemark{c}}             &
\colhead{I}                              &
\colhead{I}                              &
\colhead{I}                              &
\colhead{I}                              &
\colhead{I}                              &
\colhead{II}                             &
\colhead{II}                             &
\colhead{II}                             &
\colhead{I}                              &
\colhead{I}                              &
\colhead{I}                              &
\colhead{I}                              &
\colhead{I}                              &
\colhead{I}                              &
\colhead{II}                             &
\colhead{II}                             &
\colhead{II}                              
}
\startdata
 AS014500  &   0.46   &   0.25   &     3    &   0.05   &  \nodata &     1    &     0.47 &   0.03   &     3    &   0.48   &   0.27   &     7    &  \nodata &  \nodata &  \nodata &     0.44 &   0.12   &    15    \\   
 AS023706  &   0.43   &   0.30   &     4    &   -0.17  &  \nodata &     1    &     0.30 &   0.09   &     3    &   0.34   &   0.14   &    12    &     0.26 &   0.11   &     4    &     0.46 &   0.36   &    19    \\   
 AS081933  &   0.44   &   0.12   &     4    &   -0.03  &  \nodata &     1    &     0.89 &   0.36   &     2    &   0.28   &   0.30   &     2    &  \nodata &  \nodata &  \nodata &     0.62 &   0.12   &    10    \\   
 AS085254  &   0.67   &   0.27   &     4    &   0.41   &  \nodata &     1    &     0.26 &   0.11   &     3    &   0.30   &   0.41   &     4    &     0.12 &  \nodata &     1    &     0.51 &   0.27   &    15    \\   
 AS090900  &   0.17   &   0.37   &     3    &   0.52   &  \nodata &     1    &     0.53 &   0.39   &     2    &   0.44   &   0.23   &     2    &  \nodata &  \nodata &  \nodata &     0.37 &   0.18   &    13    \\   
 AS094541  &   0.64   &   0.36   &     4    &   0.48   &  \nodata &     1    &     0.59 &  \nodata &     1    &   0.37   &   0.21   &     7    &     0.60 &   0.13   &     2    &     0.63 &   0.22   &    13    \\   
 AS095328  &   0.51   &   0.18   &     3    &   0.09   &  \nodata &     1    &     0.51 &   0.01   &     2    &   0.40   &   0.16   &     8    &     0.33 &   0.14   &     3    &     0.50 &   0.31   &    17    \\   
 AS101332  &   0.25   &   0.38   &     3    &   0.31   &  \nodata &     1    &     0.71 &   0.35   &     2    &   0.36   &   0.14   &    11    &     0.21 &   0.22   &     4    &     0.42 &   0.17   &    15    \\   
 AS110522  &   0.49   &   0.17   &     3    &   0.29   &  \nodata &     1    &     0.19 &   0.04   &     2    &   0.60   &   0.41   &     2    &  \nodata &  \nodata &  \nodata &     0.62 &   0.29   &    14    \\   
 AS123811  &   0.61   &   0.28   &     3    &   0.36   &  \nodata &     1    &     0.25 &   0.24   &     3    &   0.32   &   0.15   &    11    &     0.41 &   0.20   &     7    &     0.42 &   0.17   &    16    \\   
 AS132225  &   0.38   &   0.09   &     3    &   0.11   &  \nodata &     1    &     0.29 &   0.13   &     3    &   0.10   &   0.16   &    13    &     0.20 &   0.19   &     3    &     0.19 &   0.23   &    13    \\   
 AS132448  &   0.81   &   0.30   &     3    &   0.84   &  \nodata &     1    &     0.40 &   0.14   &     2    &   0.52   &   0.39   &     4    &  \nodata &  \nodata &  \nodata &     0.67 &   0.38   &    14    \\   
 AS143322  &   0.52   &  \nodata &     1    &   0.50   &  \nodata &     1    &     0.51 &   0.20   &     3    &   0.35   &   0.19   &     9    &     0.27 &   0.19   &     3    &     0.36 &   0.18   &    12    \\   
 AS162158  &   0.11   &   0.26   &     3    &   0.02   &  \nodata &     1    &  \nodata &  \nodata &  \nodata &   -0.06  &  \nodata &     1    &  \nodata &  \nodata &  \nodata &     0.47 &   0.13   &     9    \\   
 AS190212  &   0.27   &   0.20   &     4    &   -0.01  &  \nodata &     1    &     0.50 &   0.12   &     2    &   0.26   &   0.07   &     2    &  \nodata &  \nodata &  \nodata &     0.60 &   0.16   &    13    \\   
 AS200431  &   0.06   &   0.12   &     3    &   -0.06  &  \nodata &     1    &     0.32 &  \nodata &     1    &  \nodata &  \nodata &  \nodata &  \nodata &  \nodata &  \nodata &     0.56 &   0.12   &    12    \\   
 AS203145  &   0.33   &   0.23   &     3    &   0.63   &  \nodata &     1    &     0.65 &   0.11   &     2    &   0.32   &   0.17   &    14    &     0.28 &   0.11   &     5    &     0.28 &   0.20   &    15    \\   
 AS211933  &   0.70   &   0.18   &     3    &   0.20   &  \nodata &     1    &     0.48 &   0.15   &     3    &   0.39   &   0.15   &    11    &     0.38 &   0.04   &     3    &     0.50 &   0.36   &    15    \\   
 AS230659  &   0.38   &  \nodata &     1    &   0.35   &  \nodata &     1    &     0.25 &   0.30   &     2    &   0.08   &   0.35   &     6    &     0.28 &   0.23   &     2    &     0.26 &   0.17   &     7    \\   
 \\      
mean     &   0.43   &          &          &   0.26   &          &          &     0.45 &          &          &   0.33   &          &          &     0.30 &          &          &     0.47 &          &          \\
$\pm$    &   0.05   &          &          &   0.06   &          &          &     0.05 &          &          &   0.04   &          &          &     0.04 &          &          &     0.03 &          &          \\   
num      &    18    &          &          &    18    &          &          &       17 &          &          &    17    &          &          &       10 &          &          &       18 &          &          \\   
$\sigma$(mean)         
         &   0.21   &          &          &   0.27   &          &          &     0.19 &          &          &   0.16   &          &          &     0.13 &          &          &     0.13 &          &          
\enddata

\tablenotetext{a}{$\sigma$ = the standard deviation of an abundance}
\tablenotetext{b}{\# = the number of lines contributing to an abundance}
\tablenotetext{c}{the I's and II's of this row are species ionization stages}

\end{deluxetable}
\end{center}

\clearpage
\begin{center}
\begin{deluxetable}{crrrrrrrrrrrrrrrrrr}
\tabletypesize{\scriptsize}
\tablewidth{0pt}
\tablecaption{Relative Abundances of Fe-group and $n$-capture Elements\label{tab-fegroup}}
\tablecolumns{19}
\tablehead{
\colhead{Star}                           &
\colhead{[Sc/Fe]}                        &
\colhead{$\sigma$\tablenotemark{a}}      &
\colhead{\#\tablenotemark{b}}            &
\colhead{[Cr/Fe]}                        &
\colhead{$\sigma$}                       &
\colhead{\#}                             &
\colhead{[Cr/Fe]}                        &
\colhead{$\sigma$}                       &
\colhead{\#}                             &
\colhead{[Zn/Fe]}                        &
\colhead{$\sigma$}                       &
\colhead{\#}                             &
\colhead{[Sr/Fe]}                        &
\colhead{$\sigma$}                       &
\colhead{\#}                             &
\colhead{[Ba/Fe]}                        &
\colhead{$\sigma$}                       &
\colhead{\#}                             \\
\colhead{}                               &
\colhead{II\tablenotemark{c}}            &
\colhead{II}                             &
\colhead{II}                             &
\colhead{I}                              &
\colhead{I}                              &
\colhead{I}                              &
\colhead{II}                             &
\colhead{II}                             &
\colhead{II}                             &
\colhead{I}                              &
\colhead{I}                              &
\colhead{I}                              &
\colhead{II}                             &
\colhead{II}                             &
\colhead{II}                             &
\colhead{II}                             &
\colhead{II}                             &
\colhead{II}
}
\startdata
AS014500 &     0.03 &     0.14 &        4 &  $-$0.07 &     0.14 &        3 &     0.14 &     0.16 &        4 &  \nodata &  \nodata &  \nodata &     0.24 &     0.06 &        2 &     0.12 &     0.02 &        2 \\ 
AS023706 &  $-$0.13 &     0.10 &        5 &  $-$0.04 &     0.11 &        4 &  $-$0.07 &     0.16 &        4 &     0.12 &  \nodata &        1 &     0.40 &  \nodata &        1 &  $-$0.03 &     0.14 &        3 \\ 
AS081933 &     0.40 &     0.26 &        4 &  $-$0.08 &  \nodata &        1 &     0.07 &  \nodata &        1 &  \nodata &  \nodata &  \nodata &     0.06 &     0.19 &        2 &  \nodata &  \nodata &  \nodata \\
AS085254 &     0.17 &     0.22 &        5 &  $-$0.10 &     0.21 &        3 &     0.01 &     0.09 &        4 &     0.25 &  \nodata &        1 &     0.10 &  \nodata &        1 &     0.56 &     0.90 &        2 \\ 
AS090900 &  $-$0.06 &     0.29 &        3 &     0.10 &     0.05 &        3 &  \nodata &  \nodata &  \nodata &  \nodata &  \nodata &  \nodata &  $-$0.17 &     0.01 &        2 &     0.19 &     0.17 &        3 \\ 
AS094541 &     0.17 &     0.06 &        4 &  \nodata &  \nodata &  \nodata &     0.11 &     0.10 &        3 &  \nodata &  \nodata &  \nodata &  \nodata &  \nodata &  \nodata &     0.13 &     0.02 &        2 \\ 
AS095328 &     0.13 &     0.26 &        4 &  $-$0.18 &     0.12 &        4 &  $-$0.08 &     0.04 &        4 &     0.30 &     0.01 &        2 &     0.70 &     0.00 &        2 &     0.16 &     0.10 &        3 \\ 
AS101332 &  $-$0.12 &     0.18 &        3 &  $-$0.03 &     0.12 &        5 &  $-$0.09 &     0.06 &        3 &     0.33 &     0.14 &        2 &  \nodata &  \nodata &  \nodata &     0.30 &     0.21 &        3 \\ 
AS110522 &     0.43 &     0.27 &        4 &  $-$0.04 &     0.18 &        4 &     0.15 &     0.16 &        3 &  \nodata &  \nodata &  \nodata &  $-$0.03 &     0.40 &        2 &  $-$0.24 &     0.05 &        2 \\ 
AS123811 &  $-$0.03 &     0.13 &        5 &  $-$0.19 &     0.12 &        4 &  $-$0.06 &     0.13 &        4 &     0.56 &  \nodata &        1 &  \nodata &  \nodata &  \nodata &  $-$0.01 &     0.29 &        3 \\ 
AS132225 &  $-$0.27 &     0.13 &        4 &  $-$0.07 &     0.22 &        4 &  $-$0.01 &     0.10 &        4 &     0.30 &  \nodata &        1 &  \nodata &  \nodata &  \nodata &  $-$0.50 &     0.31 &        3 \\ 
AS132448 &     0.32 &     0.20 &        4 &  $-$0.16 &     0.07 &        3 &     0.05 &     0.10 &        2 &  \nodata &  \nodata &  \nodata &     0.12 &     0.11 &        2 &  $-$0.32 &     0.12 &        2 \\ 
AS143322 &  $-$0.02 &     0.21 &        4 &  $-$0.08 &     0.11 &        4 &  $-$0.08 &     0.06 &        3 &     0.32 &     0.20 &        2 &  \nodata &  \nodata &  \nodata &  $-$0.16 &     0.11 &        2 \\ 
AS162158 &     0.19 &     0.24 &        3 &  $-$0.11 &     0.16 &        2 &     0.11 &     0.04 &        2 &  \nodata &  \nodata &  \nodata &  $-$0.17 &  \nodata &        1 &     0.23 &     0.15 &        2 \\ 
AS190212 &     0.30 &     0.05 &        2 &     0.00 &     0.03 &        3 &     0.27 &     0.04 &        2 &  \nodata &  \nodata &  \nodata &  $-$0.31 &     0.01 &        2 &  \nodata &  \nodata &  \nodata \\
AS200431 &     0.36 &     0.19 &        2 &     0.17 &     0.14 &        2 &     0.23 &     0.07 &        2 &  \nodata &  \nodata &  \nodata &  $-$0.81 &     0.08 &        2 &  \nodata &  \nodata &  \nodata \\
AS203145 &  $-$0.68 &     0.12 &        4 &  $-$0.09 &     0.26 &        5 &     0.06 &     0.22 &        4 &     0.32 &     0.02 &        2 &  \nodata &  \nodata &  \nodata &  $-$0.43 &     0.04 &        3 \\ 
AS211933 &  $-$0.03 &     0.29 &        4 &  $-$0.10 &     0.10 &        4 &  $-$0.01 &     0.08 &        4 &     0.25 &     0.05 &        2 &  \nodata &  \nodata &  \nodata &  $-$0.27 &     0.03 &        2 \\ 
AS230659 &     0.04 &     0.60 &        5 &  $-$0.11 &     0.18 &        3 &     0.09 &     0.14 &        4 &  \nodata &  \nodata &  \nodata &     1.13 &     0.22 &        2 &     0.64 &  \nodata &        1 \\ 
 \\
mean   &     0.06 &          &          &  $-$0.07 &          &          &     0.05 &          &          &     0.30 &          &          &     0.11 &          &          &     0.02 &          &          \\   
$\pm$  &     0.06 &          &          &     0.02 &          &          &     0.03 &          &          &     0.04 &          &          &     0.15 &          &          &     0.09 &          &          \\
num    &       18 &          &          &       17 &          &          &       17 &          &          &        9 &          &          &       11 &          &          &       15 &          &          \\   
$\sigma$(mean)       
       &     0.27 &          &          &     0.09 &          &          &     0.11 &          &          &     0.11 &          &          &     0.50 &          &          &     0.33 &          &          
\enddata

\tablenotetext{a}{$\sigma$ = the standard deviation of an abundance}
\tablenotetext{b}{\# = the number of lines contributing to an abundance}
\tablenotetext{c}{the I's and II's of this row are species ionization stages}

\end{deluxetable}
\end{center}

\clearpage
\begin{center}
\begin{deluxetable}{crrr}
\tabletypesize{\footnotesize}
\tablewidth{0pt}
\tablecaption{Radial Velocity Shift and Stretch Parameters\label{tab-shift}}
\tablecolumns{4}
\tablehead{
\colhead{Star}                                   &
\colhead{Shift\tablenotemark{a}}                 &
\colhead{$\Delta$RV=24\tablenotemark{b}}         &
\colhead{$\Delta$RV=30\tablenotemark{b}}         \\
\colhead{}                                       &
\colhead{\kmsec}                                 &
\colhead{\kmsec}                                 &
\colhead{\kmsec}                                  
}
\startdata
AS023706   &      151.0  &  0.816  &   1.02   \\
AS094541   &   $-$200.0  &  0.864  &   1.08   \\
AS123811   &   $-$147.0  &  0.938  &   1.17   \\
AS132225   &        5.4  &  0.944  &   1.18   \\
AS132448   &   $-$104.5  &  0.912  &   1.14   \\
AS190212   &    $-$87.2  &  1.112  &   1.39   \\
AS211933   &      133.3  &  0.944  &   1.18   \\
          &              &         &          \\
  mean    &              &  0.933  &   1.17   \\
$\sigma$  &              &  0.092  &   0.12   \\
\enddata

\tablenotetext{a}{the s2 parameter for $\Delta RV$ discussed 
                  in \S\ref{templates}}
\tablenotetext{b}{the s1 parameter for $\Delta RV$ discussed 
                  in \S\ref{templates}}

\end{deluxetable}
\end{center}

\clearpage
\begin{center}
\begin{deluxetable}{rrrrr}
\tabletypesize{\scriptsize}
\tablewidth{0pt}
\tablecaption{Mean Velocity Curve Templates\label{tab-curvemean}}
\tablecolumns{5}
\tablehead{
\colhead{$\phi$}                                 &
\colhead{$\Delta$RV(s1,s2)}                      &
\colhead{$\Delta$RV(s1,s2)}                      &
\colhead{$\phi$}                                 &
\colhead{$\Delta$RV(s1,s2)}                      \\
\colhead{}                                       &
\colhead{$\Delta$RV=30}                          &
\colhead{$\Delta$RV=24}                          &
\colhead{}                                       &
\colhead{$\Delta$RV=40}                          \\
\colhead{}                                       &
\colhead{\kmsec}                                 &
\colhead{\kmsec}                                 &
\colhead{}                                       &
\colhead{\kmsec}
}
\startdata
   0.00  & $-$17.5  & $-$14.0  &   0.013  & $-$18.6 \\
   0.02  & $-$17.2  & $-$13.8  &   0.047  & $-$23.1 \\   
   0.04  & $-$16.5  & $-$13.2  &   0.080  & $-$24.6 \\
   0.06  & $-$15.7  & $-$12.6  &   0.113  & $-$22.3 \\   
   0.08  & $-$14.8  & $-$11.8  &   0.147  & $-$19.8 \\
   0.10  & $-$13.9  & $-$11.1  &   0.180  & $-$17.1 \\   
   0.12  & $-$12.8  & $-$10.2  &   0.213  & $-$13.9 \\
   0.14  & $-$11.7  &  $-$9.4  &   0.246  & $-$10.5 \\   
   0.16  & $-$10.4  &  $-$8.3  &   0.280  &  $-$7.2 \\
   0.18  &  $-$9.0  &  $-$7.2  &   0.313  &  $-$4.0 \\   
   0.20  &  $-$7.5  &  $-$6.0  &   0.346  &  $-$0.9 \\
   0.22  &  $-$6.0  &  $-$4.8  &   0.380  &     2.4 \\   
   0.24  &  $-$4.4  &  $-$3.5  &   0.413  &     5.9 \\
   0.26  &  $-$3.0  &  $-$2.4  &   0.446  &     9.1 \\   
   0.28  &  $-$1.7  &  $-$1.4  &   0.480  &    11.4 \\
   0.30  &  $-$0.5  &  $-$0.4  &   0.513  &    13.2 \\   
   0.32  &     0.6  &     0.5  &   0.546  &    13.8 \\
   0.34  &     1.4  &     1.1  &   0.579  &    13.7 \\   
   0.36  &     2.2  &     1.8  &   0.613  &    13.4 \\
   0.38  &     3.1  &     2.5  &   0.646  &    13.0 \\   
   0.40  &     3.9  &     3.1  &   0.679  &    13.2 \\
   0.42  &     4.8  &     3.8  &   0.713  &    14.0 \\          
   0.44  &     5.5  &     4.4  &   0.746  &    14.6 \\
   0.46  &     6.5  &     5.2  &   0.779  &    15.4 \\          
   0.48  &     7.2  &     5.8  &   0.813  &    14.4 \\
   0.50  &     7.9  &     6.3  &   0.846  &    11.2 \\          
   0.52  &     8.7  &     7.0  &   0.879  &     4.4 \\
   0.54  &     9.4  &     7.5  &   0.912  &  $-$2.6 \\          
   0.56  &    10.1  &     8.1  &   0.946  &  $-$7.1 \\
   0.58  &    10.7  &     8.6  &   0.979  & $-$12.6 \\          
   0.60  &    11.2  &     9.0  &   1.012  & $-$18.6 \\
   0.62  &    11.7  &     9.4  &          &         \\          
   0.64  &    11.9  &     9.5  &          &         \\
   0.66  &    12.0  &     9.6  &          &         \\
   0.68  &    12.1  &     9.7  &          &         \\
   0.70  &    12.1  &     9.7  &          &         \\
   0.72  &    11.8  &     9.4  &          &         \\
   0.74  &    11.4  &     9.1  &          &         \\
   0.76  &    10.4  &     8.3  &          &         \\
   0.78  &     9.4  &     7.5  &          &         \\
   0.80  &     8.2  &     6.6  &          &         \\
   0.82  &     6.9  &     5.5  &          &         \\
   0.84  &     5.2  &     4.2  &          &         \\
   0.86  &     3.3  &     2.6  &          &         \\
   0.88  &     0.3  &     0.2  &          &         \\
   0.90  &  $-$4.0  &  $-$3.2  &          &         \\
   0.92  &  $-$8.5  &  $-$6.8  &          &         \\
   0.94  & $-$12.5  & $-$10.0  &          &         \\
   0.96  & $-$15.8  & $-$12.6  &          &         \\
   0.98  & $-$17.3  & $-$13.8  &          &         \\
\enddata

\end{deluxetable}
\end{center}

\begin{center}
\begin{deluxetable}{cccccr}
\tabletypesize{\footnotesize}
\tablewidth{0pt}
\tablecaption{Motion Parameters for Selected RRc Program Stars\label{tab-rv}}
\tablecolumns{6}
\tablehead{
\colhead{Star}                                   &
\colhead{Primary}                                & 
\colhead{Radius}                                 &
\colhead{Metal RV}                               &
\colhead{H$\alpha$ RV}                           &
\colhead{$\gamma$--velocity\tablenotemark{a}}    \\
\colhead{}                                       &
\colhead{acceleration\tablenotemark{a}}          & 
\colhead{variation\tablenotemark{a}}             &
\colhead{amplitude\tablenotemark{a}}             &
\colhead{amplitude\tablenotemark{a}}             &
\colhead{}                                       \\
\colhead{}                                       &
\colhead{km s$^{-2}$}                            & 
\colhead{R$_\odot$}                              &
\colhead{\kmsec}                                 &
\colhead{\kmsec}                                 &
\colhead{\kmsec}                                  
}
\startdata
AS014500 &   8.51 &    0.21 &  33.60 &  44.10 &   $-$17\\
AS023706 &  13.71 &    0.22&   38.08 &  45.34 &  $-$148\\
AS094541 &  10.28 &    0.23 &  35.61 &  46.12 &     184\\
AS101332 &  12.33 &    0.19 &  31.74 &  40.26 &     215\\
AS123811 &  13.53 &    0.22 &  34.30 &  42.32 &     126\\
AS132225 &  14.59 &    0.14 &  32.09 &  38.10 &      91\\
AS132448 &  11.22 &    0.23 &  35.64 &  48.02 &      91\\
AS143322 &   4.88 &    0.07 &  14.32 &  17.21 &   $-$65\\
AS190212 &   9.29 &    0.18 &  29.13 &  40.36 &      62\\
AS203145 &  11.15 &    0.19 &  29.87 &  39.31 &    $-$5\\
AS211933 &  14.83 &    0.17 &  33.42 &  43.04 &  $-$112\\
\enddata
 \tablenotetext{a}{The error bars are 
                  $\pm$\,2.5\,km s$^{-2}$ for primary acceleraions,
                  $\pm$\,0.05\,R$\odot$ for radius variation,
                  and $\pm$\,2.5\,km s$^{-1}$ for $\gamma$--velocity}
 \end{deluxetable}
\end{center}

\begin{center}
\begin{deluxetable}{crrrrr}
\tabletypesize{\footnotesize}
\tablewidth{0pt}
\tablecaption{Rotation Parameters for Blazhko Stars\label{tab-pblvrot}}
\tablecolumns{6}
\tablehead{
\colhead{Star}                                   &
\colhead{P$_{Blazhko}$\tablenotemark{a}}         &
\colhead{$V_{rot}$}                              &
\colhead{$V_{macrot}$}                           &
\colhead{sin($i$)}                               &
\colhead{$i$}                                    \\
\colhead{}                                       &
\colhead{days}                                   &
\colhead{\kmsec}                                 &
\colhead{\kmsec}                                 &
\colhead{}                                       &
\colhead{deg}                                    \\
}
\startdata
AS081933  &   8.1  &       28  &      5.8  & $<$0.204  &  $<$11.8 \\
AS085254  &  11.8  &       20  &      6.1  & $<$0.313  &  $<$18.2 \\
AS090900  &   8.5  &       27  &     14.4  & $<$0.533  &  $<$32.2 \\
AS110522  &   7.4  &       31  &      9.6  & $<$0.309  &  $<$18.0 \\
AS162158  &   8.1  &       28  &     12.2  & $<$0.430  &  $<$25.5 \\
AS200431  &  10.8  &       21  &      7.6  & $<$0.357  &  $<$20.9 \\
AS230659  &  10.2  &       22  &     12.0  & $<$0.534  &  $<$32.3 \\
\enddata

\tablenotetext{a}{\cite{szczygiel07}}
\tablenotetext{b}{(limit) = upper limit}

\end{deluxetable}
\end{center}

\end{document}